\documentclass[prd,reprint,nofootinbib,preprintnumbers,showpacs]{revtex4-2}
\usepackage{hyperref,amsmath,amssymb,graphicx,bm,bbm,multirow}
\usepackage[T1]{fontenc} 
\usepackage{empheq,mathtools}
\usepackage{cancel}

\def\tsp{\hspace{0.05em}}
\def\arXiv#1{\href{http://arxiv.org/abs/#1}{arXiv:#1}}
\def\arXiv#1#2{\href{http://arxiv.org/abs/#1}{arXiv:#1}}
\def\arXivid#1#2{\href{http://arxiv.org/abs/#1/#2}{#1/#2}}
\def\doi#1#2{\href{http://doi.org/#1}{#2}}

\begin{document}
\title{Analytic solutions of neutral hyperbolic black holes with scalar hair}
\author{Jie Ren}
\email{renjie7@mail.sysu.edu.cn}
\affiliation{School of Physics, Sun Yat-sen University, Guangzhou, 510275, China}
\date{\today}
\begin{abstract}
We find analytic solutions of hyperbolic black holes with scalar hair in anti-de Sitter (AdS) space, and they do not have spherical or planar counterparts. The system is obtained by taking a neutral limit of an Einstein-Maxwell-dilaton system whose special cases are maximal gauged supergravities, while the dilaton is kept nontrivial. There are phase transitions between these black holes and the hyperbolic Schwarzschild-AdS black hole. We discuss two AdS/CFT applications of these hyperbolic black holes. One is phase transitions of holographic R\'{e}nyi entropies, and the other is phase transitions of quantum field theories in de Sitter space. In addition, we give a C-metric solution as a generalization of the hyperbolic black holes with scalar hair.
\end{abstract}
\maketitle
\tableofcontents

\section{Introduction}
\label{sec:intro}
The AdS/CFT correspondence provides a powerful tool to study strongly coupled conformal field theories (CFTs) in a given spacetime background $\mathcal{B}_d$ in terms of a classical gravity whose boundary is conformal to $\mathcal{B}_d$ \cite{Marolf:2013ioa}. A basic example is the Schwarzschild-AdS solution
\begin{multline}
ds^2=-\left(k-\frac{2M}{r}+\frac{r^2}{L^2}\right)dt^2\\
+\left(k-\frac{2M}{r}+\frac{r^2}{L^2}\right)^{-1}dr^2+r^2\tsp d\tsp\Sigma_{2,k}^2\,,\label{eq:Schw}
\end{multline}
where $k=1$, $0$, and $-1$ are for positive, zero, and negative curvatures of the two-dimensional space $d\tsp\Sigma_{2,k}^2$, respectively. Properties of these three cases are as follows.
\begin{itemize}
\item Spherical black hole ($k=1$): The AdS boundary is $\mathbb{R}\times S^{d-1}$, so the dual CFT lives on a sphere. Black holes have a minimum temperature. There is a Hawking-Page phase transition between a black hole and a thermal gas \cite{{Hawking:1982dh,Witten:1998zw}}.
\item Planar black hole ($k=0$): The AdS boundary is $\mathbb{R}^d$, so the dual CFT lives on a Minkowski space. There is no phase transition at finite temperature. If a spatial dimensional is compactified to $S^1$, there will be a phase transition between a black hole and the AdS soliton \cite{Witten:1998zw}.
\item Hyperbolic black hole ($k=-1$): The AdS boundary is $\mathbb{R}\times \mathbb{H}^{d-1}$, so the CFT lives on a hyperboloid. There is no phase transition \cite{Emparan:1998he,Birmingham:1998nr}. The zero-mass solution is at finite temperature, and the zero temperature solution is reached by a negative mass \cite{Emparan:1999gf}. The AdS boundary is conformal to a Rindler space \cite{Emparan:1999gf} or a de Sitter space \cite{Marolf:2010tg}.
\end{itemize}

If we include a scalar field in the system, the hyperbolic black hole described by~\eqref{eq:Schw} may have a phase transition to a hyperbolic black hole with scalar hair. The IR (near-horizon) geometry of the extremal hyperbolic black hole described by~\eqref{eq:Schw} is AdS$_2\times \mathbb{H}^2$, and instability will happen when the IR Breitenlohner-Freedman (BF) bound is violated \cite{Dias:2010ma}. (As a comparison, the extremal spherical or planar black hole is the pure AdS.) Numerical solutions of hyperbolic black holes with scalar hair were obtained \cite{Dias:2010ma,Belin:2013dva,Fang:2016ehk}. An analytic solution of a hyperbolic black hole with scalar hair called the MTZ black hole was obtained in \cite{Martinez:2004nb}.

A generalization of the Schwarzschild solution is Einstein-Maxwell-dilaton (EMD) systems. There are analytic solutions in maximal gauged supergravities whose special cases are EMD systems \cite{Cvetic:1999xp}. The most notable cases in AdS$_4$ are 1-charge, 2-charge, 3-charge, and 4-charge black holes, which are summarized in Appendix~\ref{sec:STU}. The thermodynamics of black holes in STU supergravity was studied in~\cite{Cvetic:1999ne,Cai:2004pz}, and there are phase transitions. Both gauge fields and dilaton fields are in the system; if we set the gauge fields to zero, the dilaton fields will also become zero, apparently. An EMD system whose special cases intersect with STU supergravity was found in \cite{Gao:2004tu,Gao:2004tv,Gao:2005}, and more properties of the system was studied by \cite{Sheykhi:2009pf,Hendi:2010gq,Goto:2018iay}.

We find a class of analytic solutions that describe phase transitions of hyperbolic black holes. These solutions are related to supergravity and do not have spherical or planar counterparts. We observe that there are two neutral limits for charged hyperbolic black holes that are solutions to the EMD system. One neutral limit is the solution~\eqref{eq:Schw}, in which the dilaton becomes zero. The other neutral limit is a black hole with scalar hair. We show that there exist both zeroth-order and third-order phase transitions between these two hyperbolic black holes at sufficiently low temperatures.

We discuss two applications of the hyperbolic black holes in terms of the AdS/CFT correspondence. One is phase transitions of the R\'{e}nyi entropies. R\'{e}nyi entropies as a generalization of the entanglement entropy play a key role in describing the quantum entanglement. If the entangling surface is a sphere, R\'{e}nyi entropies can be calculated in terms of hyperbolic black holes \cite{Casini:2011kv,Hung:2011nu}. The parameter $n$ of R\'{e}nyi entropies $S_n$ is related to the temperature of hyperbolic black holes: larger $n$ corresponds to a lower temperature. Hyperbolic black holes developing a scalar hair imply that R\'{e}nyi entropies have a phase transition in $n$. While previous studies constructed numerical solutions to describe such a phase transition, this work provides analytic examples.

The hyperbolic black holes can also be used to study strongly coupled quantum field theories (QFTs) in de Sitter space by the AdS/CFT correspondence, because the AdS boundary of a hyperbolic black hole is conformal to a de Sitter space in the static patch \cite{Marolf:2010tg}. A hyperbolic black hole describes a QFT in de Sitter space in the static patch at a temperature that may differ from the de Sitter temperature. Since there are phase transitions between hyperbolic black holes with and without scalar hair, the dual QFTs in de Sitter space will also have phase transitions. This result may shed some light on phase transitions of QFTs in the early Universe.

This paper is organized as follows. In Sec.~\ref{sec:two}, we present the analytic solution of neutral hyperbolic black holes with scalar hair. In Sec.~\ref{sec:ads4}, we study the thermodynamics of these hyperbolic black holes and their phase transitions. In Sec.~\ref{sec:adsd}, we give higher-dimensional solutions of hyperbolic black holes with scalar hair. In Sec.~\ref{sec:appl}, we discuss two applications of hyperbolic black holes in terms of the AdS/CFT correspondence. In Sec.~\ref{sec:C-metric}, we give a C-metric solution as a generalization of the hyperbolic black holes with scalar hair. Finally, we summarize and discuss some open questions.

In Appendix~\ref{sec:HR}, we use the holographic renormalization to derive the mass. In Appendix~\ref{sec:STU}, we summarize some special cases of STU supergravity. In Appendix~\ref{sec:constraint}, we present new insights on Einstein-scalar systems to motivate the scalar potential. In Appendix~\ref{sec:planar}, we discuss some properties of planar black holes.

\section{Two neutral limits of an Einstein-Maxwell-dilaton system}
\label{sec:two}
We study the AdS$_4$ case in detail, and put higher-dimensional cases in Sec.~\ref{sec:adsd}. The action is
\begin{equation}
S=\int d^4x\sqrt{-g}\left(R-\frac14 e^{-\alpha\phi}F^2-\frac12(\partial\phi)^2-V(\phi)\right),
\label{eq:action4}
\end{equation}
where $F=dA$. The potential of the dilaton field is
\begin{multline}
V(\phi)=-\frac{2}{(1+\alpha^2)^2L^2}\Bigl[\alpha^2(3\alpha^2-1)e^{-\phi/\alpha}\\
+8\alpha^2e^{(\alpha-1/\alpha)\phi/2}+(3-\alpha^2)e^{\alpha\phi}\Bigr],\label{eq:potential4}
\end{multline}
where $\alpha$ is a parameter, and the values of $\alpha=0$, $1/\sqrt{3}$, $1$, and $\sqrt{3}$ correspond to special cases of STU supergravity. This potential was found in \cite{Gao:2004tu}. A derivation of this potential with weaker assumptions is given in Appendix~\ref{sec:constraint}, in which we explain why this potential is ``privileged''.

The three exponentials in $V(\phi)$ are ordered by their importance. The $\phi\to 0$ behavior is $V(\phi)=-6/L^2-(1/L^2)\phi^2+\cdots$, where the first term is the cosmological constant, and the second term shows that the mass of the scalar field satisfies $m^2L^2=-2$. The scaling dimension of the dual scalar operator in the CFT is $\Delta_-=1$ or $\Delta_+=2$.

The solution of the metric $g_{\mu\nu}$, gauge field $A_\mu$, and dilaton field $\phi$ is \cite{Gao:2004tu}
\begin{align}
& ds^2 =-f(r)\tsp dt^2+\frac{1}{f(r)}\tsp dr^2+U(r)\tsp d\tsp\Sigma_{2,k}^2\,,\label{eq:sol4}\\
& A =2\sqrt{\frac{bc}{1+\alpha^2}}\left(\frac{1}{r_h}-\frac{1}{r}\right)dt\,,\\
& e^{\alpha\phi}=\left(1-\frac{b}{r}\right)^\frac{2\alpha^2}{1+\alpha^2},
\end{align}
with
\begin{equation}
\begin{split}
f &=\left(k-\frac{c}{r}\right)\left(1-\frac{b}{r}\right)^\frac{1-\alpha^2}{1+\alpha^2}+\frac{r^2}{L^2}\left(1-\frac{b}{r}\right)^\frac{2\alpha^2}{1+\alpha^2},\\
U &=r^2\left(1-\frac{b}{r}\right)^\frac{2\alpha^2}{1+\alpha^2}.\label{eq:fsol4}
\end{split}
\end{equation}
The solution has two parameters $b$ and $c$ in addition to $\alpha$.

\begin{figure*}
\centering
  \includegraphics[height=0.25\textwidth]{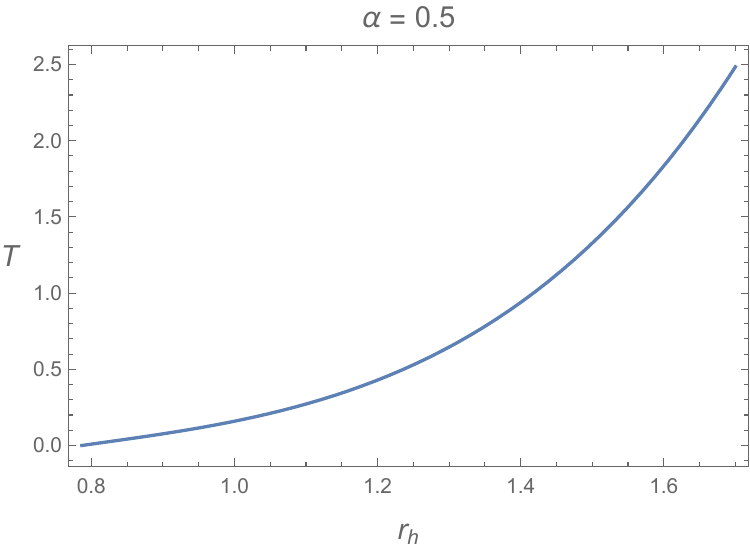}\qquad
  \includegraphics[height=0.25\textwidth]{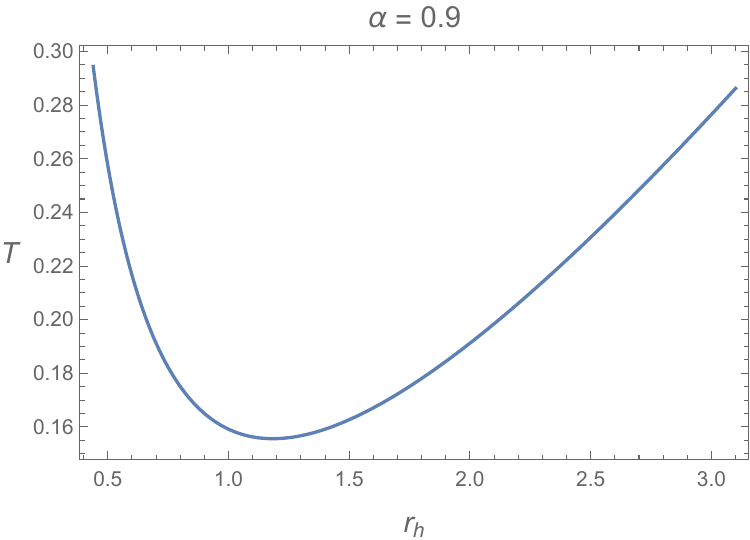}
  \caption{\label{fig:Trh4} Temperature as a function of $r_h$. In the left plot ($\alpha<1/\sqrt{3}$), there is one branch of black hole solutions, and the temperature can reach zero. In the right plot ($1/\sqrt{3}<\alpha<\sqrt{3}$), there are two branches of black hole solutions, and there is a minimum temperature.}
\end{figure*}

A key observation in this work is that we can eliminate the gauge field while keeping the dilaton field nontrivial. If we take $b=0$, this solution will be reduced to a neutral black hole described by~\eqref{eq:Schw}. If we take $c=0$, only in the case of $k=-1$ can we obtain a black hole solution. By taking $c=0$, the gauge field vanishes, but the dilaton field is still nontrivial. This is a neutral hyperbolic black hole with scalar hair. In sum, by taking a nontrivial neutral limit of the above EMD system, we find analytic solutions of hyperbolic black holes given by
\begin{equation}
ds^2=-f(r)\tsp dt^2+\frac{1}{f(r)}\tsp dr^2+U(r)\tsp d\tsp\Sigma_{2}^2\,,\label{eq:hbh}
\end{equation}
where $d\tsp\Sigma_2^2=d\theta^2+\sinh^2\theta\tsp d\varphi^2$, and
\begin{equation}
\begin{split}
f &=-\left(1-\frac{b}{r}\right)^\frac{1-\alpha^2}{1+\alpha^2}+\frac{r^2}{L^2}\left(1-\frac{b}{r}\right)^\frac{2\alpha^2}{1+\alpha^2},\\
U &=r^2\left(1-\frac{b}{r}\right)^\frac{2\alpha^2}{1+\alpha^2},\qquad e^{\alpha\phi}=\left(1-\frac{b}{r}\right)^\frac{2\alpha^2}{1+\alpha^2}.\label{eq:fsol4-1}
\end{split}
\end{equation}
Interestingly, the same type of the neutral limit of the spherical or planar black holes in the same EMD system does not give a black hole. The special case $\alpha=1/\sqrt{3}$ or $\sqrt{3}$ gives the MTZ black hole.

Another solution to the system~\eqref{eq:action4} without the gauge field is the hyperbolic Schwarzschild-AdS black hole (without scalar hair) given by
\begin{equation}
f=-1-\frac{c}{r}+\frac{r^2}{L^2}\,,\qquad U=r^2\,,\qquad \phi=0\,.\label{eq:fsol4-2}
\end{equation}
For a given $\alpha$, there is more than one black hole solution. At a given temperature, the one with lower free energy is thermodynamically preferred. In the following, we calculate thermodynamic quantities of the hyperbolic black holes with and without scalar hair, and demonstrate that there are phase transitions between the two solutions as the temperature is varied. We set the AdS radius $L=1$.

\section{Phase transitions of hyperbolic black holes}
\label{sec:ads4}

\begin{figure*}
  \includegraphics[width=0.32\textwidth]{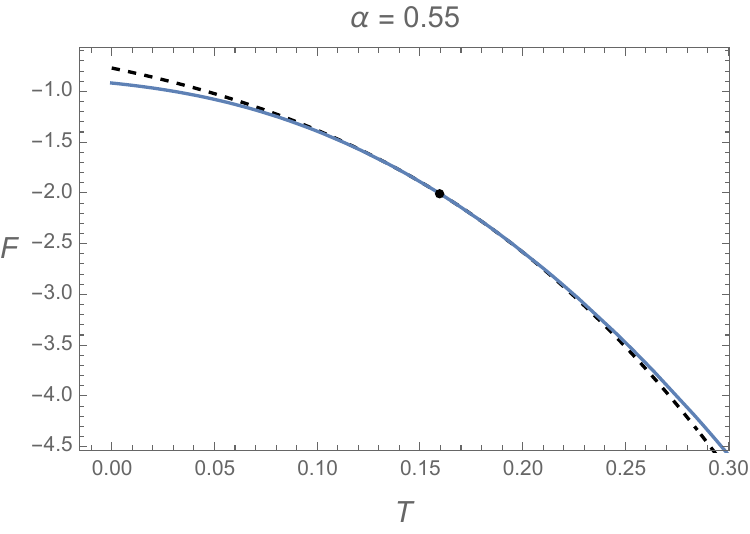}\quad
  \includegraphics[width=0.32\textwidth]{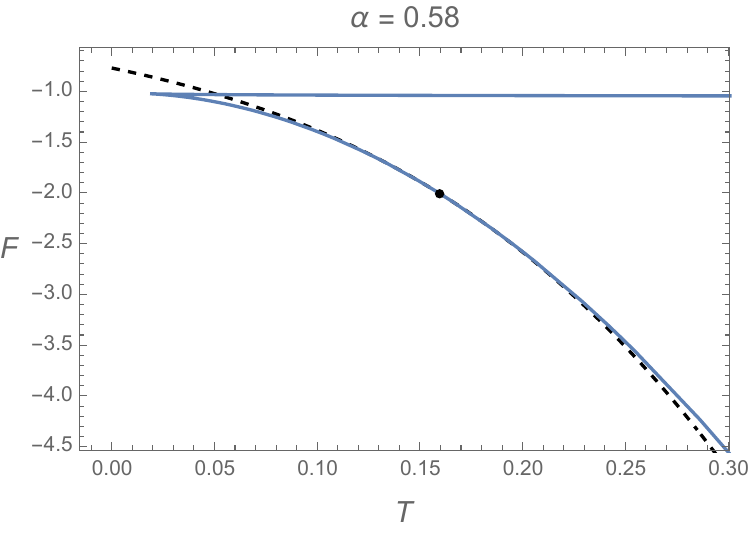}\quad
  \includegraphics[width=0.32\textwidth]{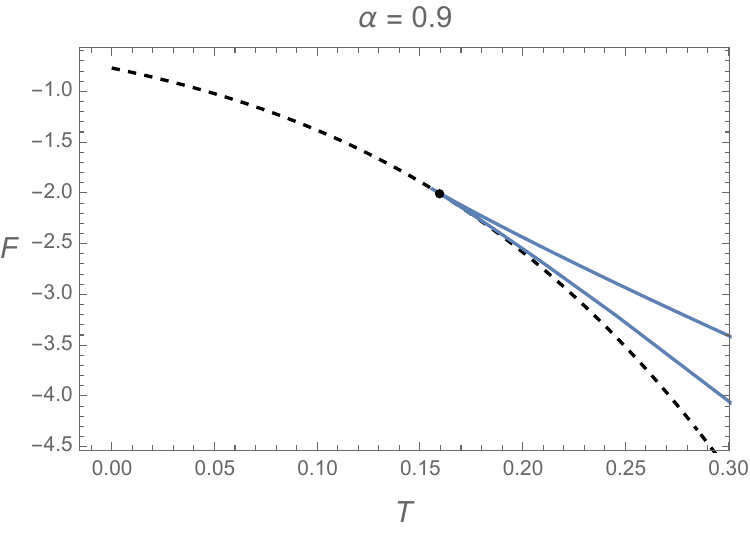}\\
  \vspace{3pt}
  \includegraphics[width=0.32\textwidth]{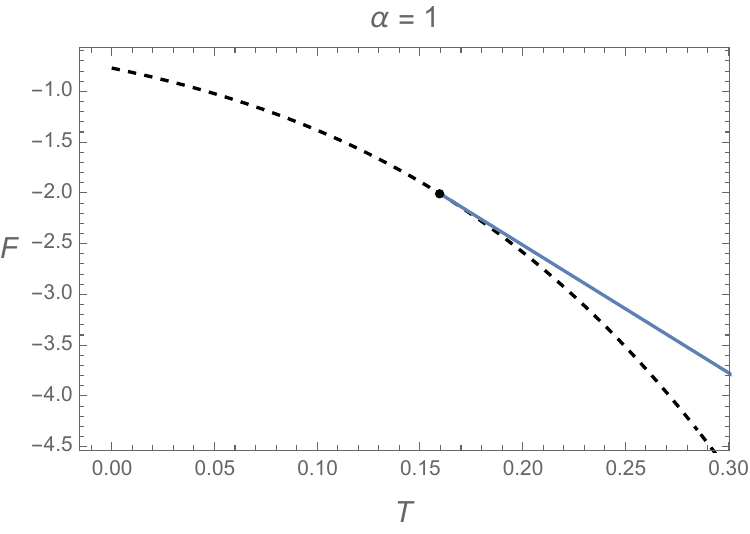}\quad
  \includegraphics[width=0.32\textwidth]{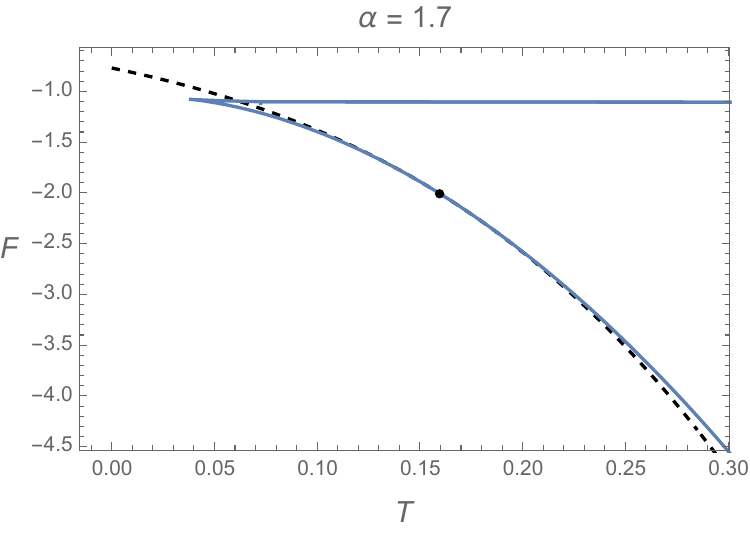}\quad
  \includegraphics[width=0.32\textwidth]{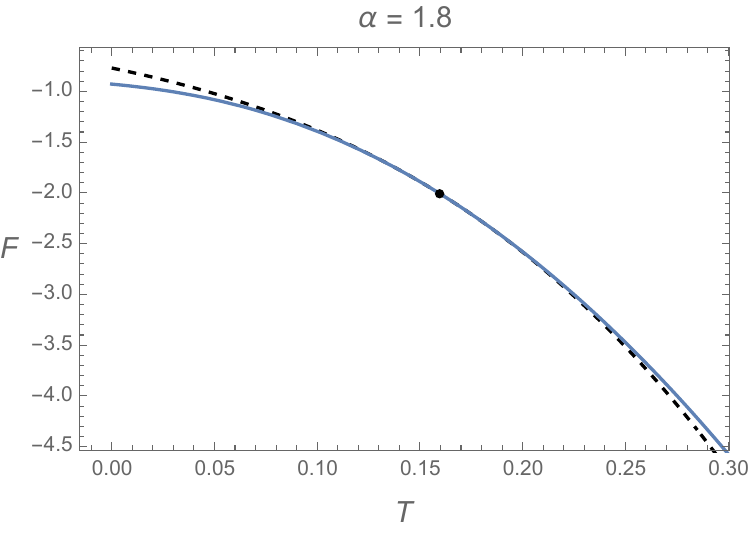}
  \caption{\label{fig:FT4} Free energy as a function of temperature for different values of $\alpha$. We take $V_\Sigma=1$. The solid line is for the hyperbolic black hole with scalar hair in AdS$_4$. The dashed line is for the hyperbolic Schwarzschild-AdS$_4$ black hole. A crossing point of the solid and the dashed lines is the pure AdS.}
\end{figure*}

Consider the hyperbolic black hole with scalar hair first. The curvature singularity is at $r=0$ and $r=b$, and the parameter $b$ can be either positive or negative. The horizon of the black hole is determined by $f(r_h)=0$, where $r_h>\max{(0,b)}$. From $f(r_h)=0$, the parameter $b$ is expressed in terms of $r_h$:
\begin{equation}
b=r_h-r_h^\frac{3-\alpha^2}{1-3\alpha^2}.\label{eq:brh}
\end{equation}
The temperature is given by
\begin{align}
T &=\frac{f'(r_h)}{4\pi}\label{eq:thermo-T}\\
&=\frac{1}{4\pi(1+\alpha^2)}\biggl[(3-\alpha^2)r_h^\frac{1+\alpha^2}{1-3\alpha^2}-(1-3\alpha^2)r_h^{-\frac{1+\alpha^2}{1-3\alpha^2}}\biggr],\nonumber
\end{align}
where we have used~\eqref{eq:brh} to replace $b$ with $r_h$. In the exceptional case $\alpha=1/\sqrt{3}$, we have $r_h=1$, and the temperature is $T=\sqrt{1-b}/2\pi$. The system is invariant under $\alpha\to-\alpha$ and $\phi\to-\phi$. We assume $\alpha>0$, and there are two distinctive cases as follows. See Fig.~\ref{fig:Trh4}.

\begin{itemize}
\item $\alpha<1/\sqrt{3}$ or $\alpha>\sqrt{3}$. The temperature reaches zero when
\begin{equation}
r_h=\left(\frac{1-3\alpha^2}{3-\alpha^2}\right)^\frac{1-3\alpha^2}{2(1+\alpha^2)}.
\end{equation}
The extremal geometry is given by~\eqref{eq:hbh} with
\begin{equation}
b=\frac{2(1+\alpha^2)}{3-\alpha^2}\left(\frac{1-3\alpha^2}{3-\alpha^2}\right)^\frac{1-3\alpha^2}{2(1+\alpha^2)}.
\end{equation}

\item $1/\sqrt{3}<\alpha<\sqrt{3}$. The temperature can never reach zero. There is a minimum temperature at
\begin{equation}
r_h=\left(\frac{3\alpha^2-1}{3-\alpha^2}\right)^\frac{1-3\alpha^2}{2(1+\alpha^2)}.
\end{equation}
For a given temperature above the minimum temperature, there are two values of $r_h$. (When $\alpha=1$, the two values coincide.) At the minimum temperature, the geometry is given by~\eqref{eq:hbh} with
\begin{equation}
b=\frac{4(1-\alpha^2)}{3-\alpha^2}\left(\frac{3\alpha^2-1}{3-\alpha^2}\right)^\frac{1-3\alpha^2}{2(1+\alpha^2)}.
\end{equation}
\end{itemize}

To show that there is a phase transition between the black holes with and without scalar hair in the canonical ensemble, we need to compare their free energies as a function of temperature. Below a critical temperature, the hairy black hole has lower free energy. Moreover, we need to impose a sourceless boundary condition for the scalar field at the AdS boundary. The asymptotic behavior of the scalar field near the AdS boundary $r\to\infty$ implies that there is a multi-trace deformation in the dual CFT \cite{Witten:2001ua}.

To specify the boundary conditions clearly, we use the Fefferman-Graham (FG) coordinates near the AdS boundary. For details, see Appendix~\ref{sec:HR} or \cite{Caldarelli:2016nni}. The asymptotic expansion of the scalar field is
\begin{equation}
\phi=A z+B z^2+\cdots,
\end{equation}
where $z$ is the FG radial coordinate as in \eqref{eq:FG} below. Starting with the alternative quantization, in which the scaling dimension of the dual scalar operator is $\Delta_-=1$, we require that the single-trace source is zero. The boundary condition compatible with the solution~\eqref{eq:fsol4-1} can be chosen as follows:
\begin{itemize}
\item $\displaystyle B/A=\frac{1-\alpha^2}{2(1+\alpha^2)}b$. This corresponds to a double-trace deformation, which is relevant.
\item $\displaystyle B/A^2=-\frac{1-\alpha^2}{4\alpha}$, which is dimensionless. This corresponds to a triple-trace deformation, which is marginal.
\end{itemize}
We use the triple-trace deformation, for which we can vary the temperature with a fixed $B/A^2$.

The mass calculated by holographic renormalization is\footnote{The mass and entropy in arbitrary dimensions are~\eqref{eq:thermo-Md} and~\eqref{eq:thermo-Sd} in Sec.~\ref{sec:adsd}. They depend on Newton's constant $G$. Here we have set $16\pi G=1$ in the action.}
\begin{equation}
M=-\frac{V_\Sigma}{8\pi G}\frac{1-\alpha^2}{1+\alpha^2}b\,.\label{eq:thermo-M}
\end{equation}
where $V_\Sigma$ is the area of the hyperbolic space $d\Sigma_2^2$, which needs to be regulated. The entropy is
\begin{equation}
S =\frac{V_\Sigma}{4G}U(r_h)=\frac{V_\Sigma}{4G}r_h^\frac{2(1-\alpha^2)}{1-3\alpha^2}.\label{eq:thermo-S}
\end{equation}
It can be checked that the first law of thermodynamics $dM=TdS$ is satisfied by~\eqref{eq:thermo-T}, \eqref{eq:thermo-M}, and \eqref{eq:thermo-S}. The free energy is
\begin{equation}
F=M-TS=-\frac{V_\Sigma}{16\pi G}\Bigl(r_h+r_h^\frac{3-\alpha^2}{1-3\alpha^2}\Bigr)\,.
\label{eq:thermo-F}
\end{equation}
The free energy as a function of temperature is plotted in Fig.~\ref{fig:FT4}.

\begin{table*}
\caption{\label{tab:FT-4} Thermodynamic quantities of the AdS$_4$ system in special cases.}
\renewcommand{\arraystretch}{2.5}
\setlength\doublerulesep{0.2pt}
\begin{ruledtabular}
\begin{tabular}{ccccccc}
& $\alpha$ & $b$ & $T$ & $S$ & $F$ & $F(T)$\\
\hline
1-charge BH & $\sqrt{3}$ & $r_h-1$ & $\dfrac{\sqrt{r_h}}{2\pi}$ & $\dfrac{V_\Sigma}{4G}\sqrt{r_h}$ & $-\dfrac{V_\Sigma}{16\pi G}(r_h+1)$ & $-\dfrac{V_\Sigma}{16\pi G}(1+4\pi^2T^2)$\\
2-charge BH & $1$ & $r_h-r_h^{-1}$ & $\dfrac{r_h^2+1}{4\pi r_h}$ & $\dfrac{V_\Sigma}{4G}$ & $-\dfrac{V_\Sigma}{16\pi G}\dfrac{r_h^2+1}{r_h}$ & $-\dfrac{V_\Sigma}{4G}\,T$\\
3-charge BH & $1/\sqrt{3}$ & $(r_h=1)$ & $\dfrac{\sqrt{1-b}}{2\pi}$ & $\dfrac{V_\Sigma}{4G}\sqrt{1-b}$ & $-\dfrac{V_\Sigma}{16\pi G}(2-b)$ & $-\dfrac{V_\Sigma}{16\pi G}(1+4\pi^2T^2)$\\
SAdS BH & $0$ & $r_h-r_h^3$ & $\dfrac{3r_h^2-1}{4\pi r_h}$ & $\dfrac{V_\Sigma}{4G}r_h^2$ & $-\dfrac{V_\Sigma}{16\pi G}r_h(r_h^2+1)$ &
\end{tabular}
\end{ruledtabular}
\end{table*}

For the hyperbolic Schwarzschild-AdS$_4$ black hole, the thermodynamic quantities can be calculated by setting $\alpha=0$ and $b=-c$. They are
\begin{equation}
\bar{T}=\frac{3\bar{r}_h^2-1}{4\pi r_h}\,,\qquad \bar{S}=\frac{V_\Sigma}{4G}\bar{r}_h^2\,,\qquad \bar{M}=\frac{V_\Sigma}{8\pi G}\bar{r}_h(\bar{r}_h^2-1)\,.
\end{equation}
The free energy is $\bar{F}=\bar{M}-\bar{T}\bar{S}$. The free energy as a function of temperature is plotted in Fig.~\ref{fig:FT4}. The solid line is for the black hole with scalar hair, and the dashed line is for the black hole without scalar hair. The two solutions~\eqref{eq:fsol4-1} and \eqref{eq:fsol4-2} share the same geometry when $b=c=0$, which is the pure AdS. This is at $r_h=\bar{r}_h=1$, which gives
\begin{equation}
T_c=\frac{1}{2\pi}\,,\qquad F_c=-\frac{V_\Sigma}{8\pi G}\,.
\end{equation}
The two curves cross at this point. We can analytically check that $d^2F/dT^2$ is continuous, and $d^3F/dT^3$ is discontinuous at $T=T_c$. Therefore, there is a third-order phase transition at $T_c$. In addition, when $1/\sqrt{3}<\alpha<\sqrt{3}$ ($\alpha\neq 1$), the free energy is discontinuous at the minimum temperature $T_\text{min}$, and thus, there is a zeroth-order phase transition at $T=T_\text{min}$.

\begin{figure}
\centering
  \includegraphics[width=0.4\textwidth]{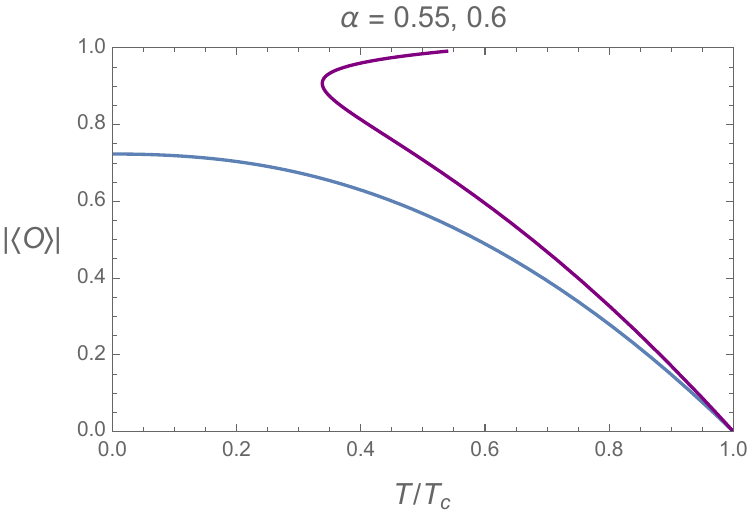}
  \caption{\label{fig:OT4} Condensation of the order parameter as a function of temperature. The blue line (lower one) is for $\alpha=0.55$, for which a third-order phase transition happens at $T=T_c$. The purple line (upper one) is for $\alpha=0.6$, for which a third-order phase transition happens at $T_c$, and a zeroth-order phase transition happens at a lower temperature $T_\text{min}$.}
\end{figure}

As the order parameter, the expectation value of the scalar operator dual to $\phi$ is $\langle\mathcal{O}\rangle=A$. Near $T_c$, we have
\begin{equation}
\langle\mathcal{O}\rangle\simeq \frac{4\pi\alpha}{1-\alpha^2}(T-T_c)\,.
\end{equation}
Figure~\ref{fig:OT4} shows the order parameter as a function of temperature.

For special values of $\alpha$ in supergravity, thermodynamic quantities of the hyperbolic black holes in AdS$_4$ are summarized in Table~\ref{tab:FT-4}. These cases have distinctive features. For $\alpha=1/\sqrt{3}$ and $\alpha=\sqrt{3}$, the entropy is linear in temperature, and the IR of the extremal geometry is conformal to AdS$_2\times \mathbb{H}^2$. The hairy black holes have lower free energy at $0\leq T<T_c$. For $\alpha=1$, the entropy is independent of temperature, and the minimum temperature equals $T_c$.

\section{Higher-dimensional solutions}
\label{sec:adsd}

The above result can be generalized to higher-dimensional spacetimes. The $(d+1)$-dimensional action is
\begin{equation}
S=\int d^{d+1}x\sqrt{-g}\biggl(R-\frac14 e^{-\alpha\phi}F^2-\frac12(\partial\phi)^2-V(\phi)\biggr),\label{eq:actiond}
\end{equation}
where $F=dA$. The potential of the dilaton field is given by
\begin{equation}
V(\phi)=v_1e^{-\frac{2(d-2)}{(d-1)\alpha}\phi} +v_2e^{\frac{(d-1)\alpha^2-2(d-2)}{2(d-1)\alpha}\phi}+v_3e^{\alpha  \phi }\,,\label{eq:potentiald}
\end{equation}
where
\begin{equation}
\begin{split}
&v_1=-\frac{(d-1)^2[d(d-1)\alpha^2-2(d-2)^2]\alpha^2}{L^2[2(d-2)+(d-1)\alpha ^2]^2}\,,\\
&v_2=-\frac{8(d-2)(d-1)^3\alpha^2}{{L^2[2(d-2)+(d-1)\alpha^2]^2}}\,,\\
&v_3=-\frac{2(d-2)^2(d-1)[2d-(d-1)\alpha^2]}{L^2[2(d-2)+(d-1)\alpha^2]^2}\,.
\end{split}
\end{equation}
The $\phi\to 0$ behavior is
\begin{equation}
V(\phi)=-\frac{d(d-1)}{L^2}-\frac{d-2}{L^2}\phi^2+\mathcal{O}(\phi^3)\,,
\end{equation}
where the first term is the cosmological constant, and the second term shows that the mass of the scalar field satisfies $m^2L^2=-2(d-2)$. Recall that the scaling dimension of the dual scalar operator satisfies $\Delta(\Delta-d)=m^2L^2$, and the BF bound is $m_\text{BF}^2L^2=-d^2/4$. The mass is above the BF bound for all $d$ except $d=4$, in which case the mass saturates the BF bound. The two solutions of $\Delta$ are $\Delta_\pm=2$, $d-2$.

The solutions and their thermodynamic quantities were obtained in \cite{Gao:2004tu,Gao:2004tv,Gao:2005,Sheykhi:2009pf,Hendi:2010gq,Goto:2018iay}. We observe that neutral hyperbolic black holes can be obtained by taking a neutral limit of the EMD systems while the dilaton field is kept nontrivial. We put a backslash on the parameter $c$ to indicate that it will be set to zero in this neutral limit. The solution is
\begin{align}
& ds^2 =-f(r)\tsp dt^2+\frac{1}{g(r)}\tsp dr^2+U(r)\tsp d\tsp\Sigma_{d-1,\tsp k}^2\,,\label{eq:ansatzd-1}\\
& A=2\sqrt{\frac{(d-1)\tsp b\tsp\bcancel{c}}{2(d-2)+(d-1)\alpha^2}}\biggl(\frac{1}{r_h^{d-2}}-\frac{1}{r^{d-2}}\biggr)dt\,,\\
& e^{\alpha\phi} =\biggl(1-\frac{b}{r^{d-2}}\biggr)^\frac{2(d-1)\alpha^2}{2(d-2)+(d-1)\alpha^2},
\end{align}
with
\begin{align}
f &=\biggl(k-\frac{\bcancel{c}}{r^{d-2}}\biggr)\biggl(1-\frac{b}{r^{d-2}}\biggr)^\frac{2(d-2)-(d-1)\alpha^2}{2(d-2)+(d-1)\alpha^2}\nonumber\\
&\hspace{60pt}+\frac{r^2}{L^2}\biggl(1-\frac{b}{r^{d-2}}\biggr)^\frac{2(d-1)\alpha^2}{(d-2)[2(d-2)+(d-1)\alpha^2]},\nonumber\\
g &=f(r)\biggl(1-\frac{b}{r^{d-2}}\biggr)^\frac{2(d-3)(d-1)\alpha^2}{(d-2)[2(d-2)+(d-1)\alpha^2]},\label{eq:fsold-1}\\
U &=r^2\biggl(1-\frac{b}{r^{d-2}}\biggr)^\frac{2(d-1)\alpha^2}{(d-2)[2(d-2)+(d-1)\alpha^2]}.\nonumber
\end{align}
We take $k=-1$ and $c=0$ for neutral hyperbolic black holes. We set $L=1$ in the following. 

Consider the hyperbolic black hole with scalar hair first. The curvature singularity is at $r=0$ and $r^{d-2}=b$, and the parameter $b$ can be either positive or negative. The horizon of the black hole is determined by $f(r_h)=0$, from which the parameter $b$ is expressed in terms of $r_h$:
\begin{equation}
b=r_h^{d-2}-r_h^{\frac{(d-2)^2 [2 d-(d-1) \alpha ^2]}{2 (d-2)^2-d (d-1) \alpha ^2}}.\label{eq:brhd}
\end{equation}
The temperature is given by
\begin{align}
&T=\frac{\sqrt{f'g'}}{4\pi}\biggr|_{r=r_h}=\label{eq:thermo-Td}\\
&\frac{(d-2)[2 d-(d-1) \alpha ^2] r_h^p-[2 (d-2)^2-d (d-1) \alpha ^2] r_h^{-p}}{4 \pi [2 (d-2)+(d-1) \alpha ^2]},\nonumber
\end{align}
where we have used~\eqref{eq:brhd} to replace $b$ with $r_h$, and
\begin{equation}
p=\frac{(d-2)[2(d-2)+(d-1)\alpha^2]}{2(d-2)^2-d(d-1)\alpha^2}.
\end{equation}
In the exceptional case $\alpha=(d-2)\sqrt{\frac{2}{d(d-1)}}$, we have $r_h=1$, and the temperature is $T=\sqrt{1-b}/2\pi$. The system is invariant under $\alpha\to-\alpha$ and $\phi\to-\phi$. We assume $\alpha>0$, and there are two distinctive cases as follows:
\begin{itemize}
\item $0<\alpha<(d-2)\sqrt{\frac{2}{d(d-1)}}$ or $\alpha>\sqrt{\frac{2d}{d-1}}$. The temperature reaches zero when
\begin{equation}
r_h=\biggl(\frac{2 (d-2)^2-d (d-1) \alpha ^2}{(d-2) [2 d-(d-1) \alpha ^2]}\biggr)^{\frac{1}{2 p}}.
\end{equation}

\item $(d-2)\sqrt{\frac{2}{d(d-1)}}<\alpha<\sqrt{\frac{2d}{d-1}}$. The temperature can never reach zero. There is a minimum temperature at
\begin{equation}
r_h=\biggl(\frac{d (d-1) \alpha ^2-2 (d-2)^2}{(d-2) [2 d-(d-1) \alpha ^2]}\biggr)^{\frac{1}{2 p}}.
\end{equation}
For a given temperature above the minimum temperature, there are two values of $r_h$. (When $\alpha=\sqrt{\frac{2(d-2)}{d-1}}$, the two values coincide.)
\end{itemize}

\begin{figure*}
  \includegraphics[width=0.32\textwidth]{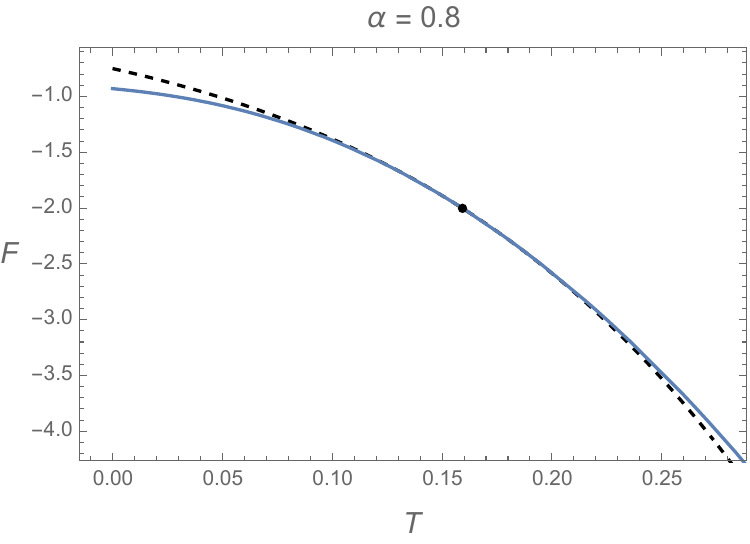}\quad
  \includegraphics[width=0.32\textwidth]{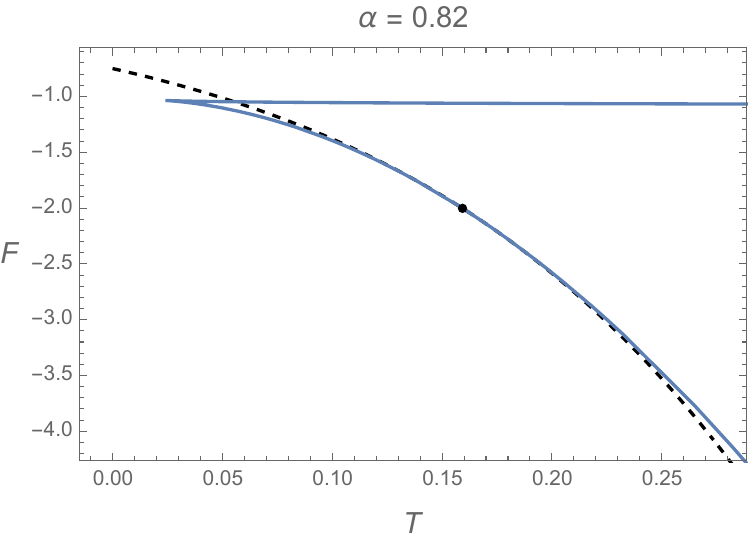}\quad
  \includegraphics[width=0.32\textwidth]{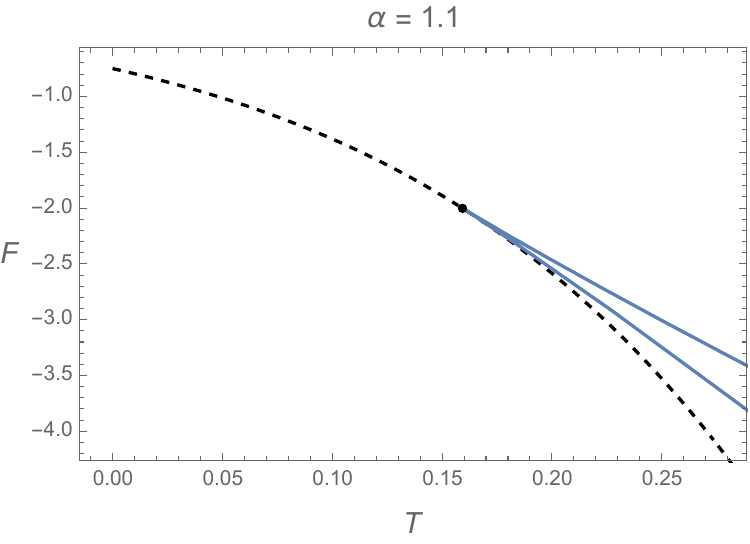}\\
  \vspace{3pt}
  \includegraphics[width=0.32\textwidth]{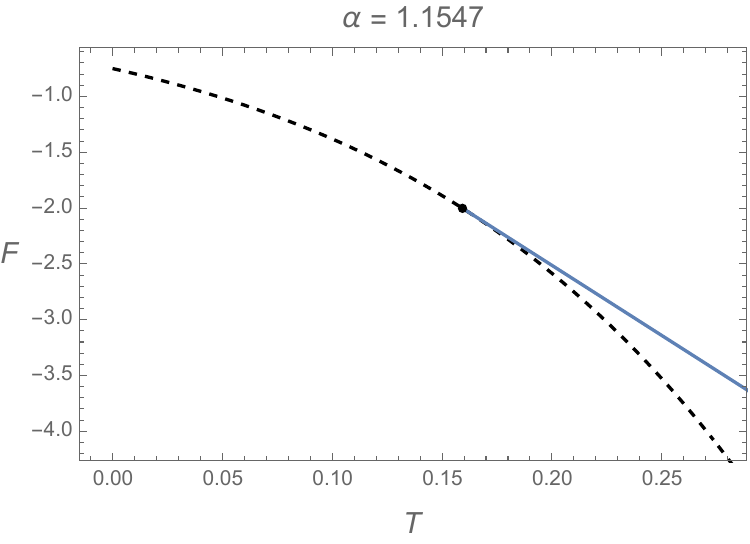}\quad
  \includegraphics[width=0.32\textwidth]{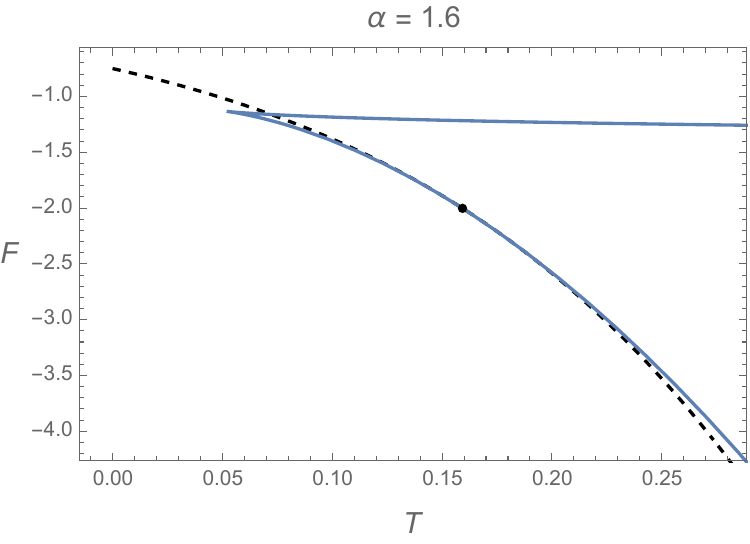}\quad
  \includegraphics[width=0.32\textwidth]{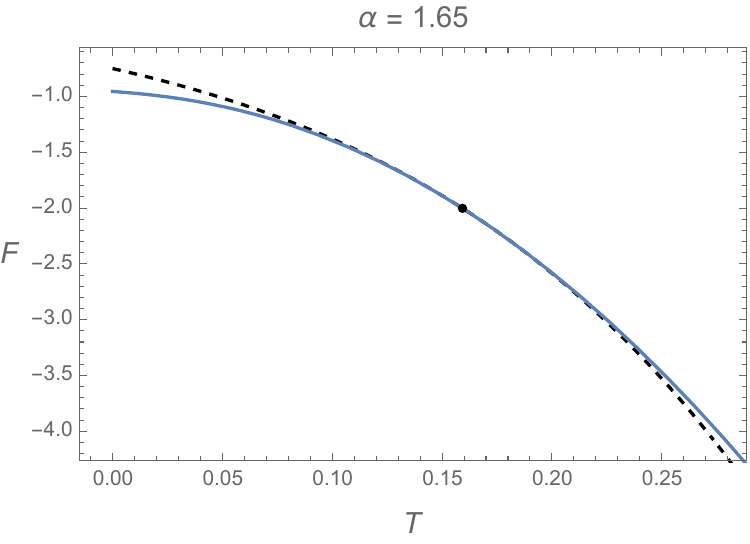}
  \caption{\label{fig:FT5} Free energy as a function of temperature for different values of $\alpha$. We take $V_\Sigma=1$. The solid line is for the hyperbolic black hole with scalar hair in AdS$_5$. The dashed line is for the hyperbolic Schwarzschild-AdS$_5$ black hole. A crossing point of the solid and the dashed lines is the pure AdS.}
\end{figure*}

The mass and entropy are \cite{Hendi:2010gq}
\begin{align}
M &=\frac{(d-1)V_\Sigma}{16\pi G}\biggl(\bcancel{c}+k\,\frac{2(d-2)-(d-1)\alpha^2}{2(d-2)+(d-1)\alpha^2}\,b\biggr),\label{eq:thermo-Md}\\
S &=\frac{V_\Sigma}{4G}U(r_h)^{(d-1)/2}\,,\label{eq:thermo-Sd}
\end{align}
where $V_\Sigma$ is the volume of the $(d-1)$-dimensional hyperbolic space, regulated by integrating out to a maximum radius in this hyperbolic geometry \cite{Hung:2011nu}:
\begin{equation}
V_\Sigma\simeq\frac{\Omega_{d-2}}{d-2}\biggl[\frac{R^{d-2}}{\delta^{d-2}}-\frac{(d-2)(d-3)}{2(d-4)}\frac{R^{d-4}}{\delta^{d-4}}+\cdots\biggr],
\end{equation}
where $\Omega_{d-2}=2\pi^{(d-1)/2}/\Gamma((d-2)/2)$ is the area of a unit $(d-2)$-sphere. The cutoff $\delta$ is related to the UV cutoff in the dual CFT, consistent with the area law of the entanglement entropy.

It can be checked that the first law of thermodynamics $dM=TdS$ is satisfied by~\eqref{eq:thermo-Td}, \eqref{eq:thermo-Md}, and \eqref{eq:thermo-Sd}. The free energy is
\begin{equation}
F=M-TS=-\frac{V_\Sigma}{16\pi G}\Bigl(r_h^{d-2}+r_h^{\frac{(d-2)^2 [2 d-(d-1) \alpha ^2]}{2 (d-2)^2-d (d-1) \alpha ^2}}\Bigr)\,.
\end{equation}

Another solution to the system~\eqref{eq:actiond} without the gauge field is the hyperbolic Schwarzschild-AdS black hole (without scalar hair) given by
\begin{equation}
f=-1-\frac{c}{r^{d-2}}+\frac{r^2}{L^2}\,,\qquad g=f\,,\qquad U=r^2\,,\qquad \phi=0\,.\label{eq:fsold-2}
\end{equation}
For a given $\alpha$, there is more than one black hole solution. The one with lower free energy is thermodynamically preferred. We demonstrate that there are phase transitions between the the solutions with and without scalar hair as the temperature is varied.

For the hyperbolic Schwarzschild-AdS$_{d+1}$ black hole, the thermodynamic quantities can be calculated by setting $\alpha=0$ and $b=-c$. They are
\begin{equation}
\begin{split}
& \bar{T}=\frac{d\bar{r}_h^2-(d-2)}{4\pi r_h}\,,\qquad \bar{S}=\frac{V_\Sigma}{4G}\bar{r}_h^{d-1}\,,\\
& \bar{M}=\frac{(d-1)V_\Sigma}{16\pi G}\bar{r}_h^{d-2}(\bar{r}_h^2-1)\,.
\end{split}
\end{equation}
The two solutions~\eqref{eq:fsold-1} and~\eqref{eq:fsold-2} share the same geometry when $b=c=0$, which is the pure AdS. This is at $r_h=\bar{r}_h=1$, which gives
\begin{equation}
T_c=\frac{1}{2\pi}\,,\qquad F_c=-\frac{V_\Sigma}{8\pi G}\,.
\end{equation}
There is a third-order phase transition at $T_c$. In addition, there is a zeroth-order phase transition at the minimum temperature $T_\text{min}$ when $(d-2)\sqrt{\frac{2}{d(d-1)}}<\alpha<\sqrt{\frac{2d}{d-1}}$ ($\alpha\neq \sqrt{\frac{2(d-2)}{d-1}}$).

As the order parameter, the expectation value of the scalar operator dual to $\phi$ is $\langle\mathcal{O}\rangle$ as in
\begin{equation}
\phi=\frac{\langle\mathcal{O}\rangle}{r^{d-2}}+\cdots.
\end{equation}
Near $T_c$, we have
\begin{equation}
\langle\mathcal{O}\rangle\simeq \frac{8\pi\alpha}{2(d-2)-(d-1)\alpha^2}(T-T_c)\,.
\end{equation}

We take a closer look at the AdS$_5$ ($d=4$) case. The hairy black hole solution is given by \eqref{eq:ansatzd-1}--\eqref{eq:fsold-1} with $d=4$, $k=-1$, and $c=0$. The temperature as a function of $r_h$ has two distinctive cases as follows: (i) $\alpha<2/\sqrt{6}$ or $\alpha>4/\sqrt{6}$: The temperature can reach zero. (ii) $2/\sqrt{6}<\alpha<4/\sqrt{6}$: The temperature can never reach zero, and there is a minimum temperature. The free energy as a function of temperature is plotted in Fig.~\ref{fig:FT5}, which is qualitatively similar to the AdS$_4$ case. The asymptotic behavior of the scalar field near the AdS boundary $r\to\infty$ is
\begin{equation}
\phi=\frac{A}{r^2}\ln r+\frac{B}{r^2}+\cdots.
\end{equation}
The boundary condition is given by the standard quantization. The source $A$ is zero, and the expectation value $B$ is the order parameter.

For special values of $\alpha$ in supergravity, thermodynamic quantities of the hyperbolic black holes in AdS$_5$ are summarized in Table~\ref{tab:FT-5}. These cases have distinctive features. For $\alpha=2/\sqrt{6}$ and $\alpha=4/\sqrt{6}$, the entropy is linear in temperature, and the IR of the extremal geometry is conformal to AdS$_2\times \mathbb{H}^3$. The hairy black holes have lower free energy at $0\leq T<T_c$.

\begin{table*}
\caption{\label{tab:FT-5} Thermodynamic quantities of the AdS$_5$ system in special cases.}
\renewcommand{\arraystretch}{2.5}
\setlength\doublerulesep{0.2pt}
\begin{ruledtabular}
\begin{tabular}{ccccccc}
& $\alpha$ & $b$ & $T$ & $S$ & $F$ & $F(T)$\\
\hline
1-charge BH & $4/\sqrt{6}$ & $r_h^2-1$ & $\dfrac{r_h}{2\pi}$ & $\dfrac{V_\Sigma}{4G}r_h$ & $-\dfrac{V_\Sigma}{16\pi G}(r_h^2+1)$ & $-\dfrac{V_\Sigma}{16\pi G}(1+4\pi^2T^2)$\\
2-charge BH & $2/\sqrt{6}$ & $(r_h=1)$ & $\dfrac{\sqrt{1-b}}{2\pi}$ & $\dfrac{V_\Sigma}{4G}\sqrt{1-b}$ & $-\dfrac{V_\Sigma}{16\pi G}(2-b)$ & $-\dfrac{V_\Sigma}{16\pi G}(1+4\pi^2T^2)$\\
SAdS BH & $0$ & $r_h^2-r_h^4$ & $\dfrac{2r_h^2-1}{2\pi r_h}$ & $\dfrac{V_\Sigma}{4G}r_h^3$ & $-\dfrac{V_\Sigma}{16\pi G}r_h^2(r_h^2+1)$ &
\end{tabular}
\end{ruledtabular}
\end{table*}

\section{AdS/CFT applications}
\label{sec:appl}
\subsection{Phase transitions of R\'{e}nyi entropies}
\label{sec:Renyi}
Consider a QFT in a state described by a density matrix $\rho$, and divide the system into two parts, A and B. The reduced density matrix for the subsystem A is $\rho_A=\text{Tr}_B\rho$. The $n$th R\'{e}nyi entropy is defined by
\begin{equation}
S_n=\frac{1}{1-n}\log\textrm{Tr}(\rho_A^n)\,.
\end{equation}
The entanglement entropy $S_{EE}$ can be obtained by taking the $n\to 1$ limit of the R\'{e}nyi entropy: $S_{EE}=\lim_{n\to 1}S_n=-\text{Tr}\rho_A\log\rho_A$. R\'{e}nyi entropies are usually difficult to calculate in QFTs.

In terms of the AdS/CFT correspondence, R\'{e}nyi entropies with the entangling surface being a sphere can be calculated by hyperbolic black holes \cite{Casini:2011kv,Hung:2011nu}. Suppose we want to calculate the R\'{e}nyi entropies of a CFT with a gravity dual, and the entangling surface between A and B is a sphere of radius $R$. By a conformal mapping, the R\'{e}nyi entropy is related to the free energy of a hyperbolic black hole:
\begin{align}
S_n &=\frac{n}{1-n}\frac{1}{T_0}(F(T_0)-F(T_0/n))\nonumber\\
&=\frac{n}{1-n}\frac{1}{T_0}\int_{T_0/n}^{T_0}S_\text{therm}(T)dT,
\label{eq:SnF}
\end{align}
where $T_0=\frac{1}{2\pi R}$ is the temperature of a zero-mass hyperbolic black hole, $S_\text{therm}$ is the thermal entropy of the hyperbolic black hole, and $S_\text{therm}=-\partial F/\partial T$. In Secs.~\ref{sec:two}--\ref{sec:adsd}, we have set $R=1$.

A phase transition of hyperbolic black holes at a sufficiently low temperature implies a phase transition of R\'{e}nyi entropies in $n$; i.e., $S_n$ is nonanalytic at some $n=n_c$. As shown in Fig.~\ref{fig:FT4}, there is always a phase transition at $T_c=1/2\pi$, i.e., $n=1$ (except for $\alpha=1$). Since the phase transition is third order, $\partial_nS_n$ is continuous and the R\'{e}nyi entropies give a well-defined entanglement entropy in the $n\to 1$ limit; see also \cite{Belin:2014mva}. For $1/\sqrt{3}<\alpha<\sqrt{3}$ ($\alpha\neq 1$), there is a minimum temperature $T_\text{min}$ for the hairy black hole, and there is a zeroth-order phase transition at $T=T_\text{min}$, corresponding to another $n_c$ of R\'{e}nyi entropies.

For general $\alpha$, the explicit $S_n$ as a function of $n$ is complicated. However, $S_n$ is strikingly simple for the dilatonic systems in supergravity. From Tables~\ref{tab:FT-4} and \ref{tab:FT-5}, we can see that the free energy as a function of temperature is $F=-\frac{V_\Sigma}{16\pi G}(1+4\pi^2T^2)$ for the $\alpha=1/\sqrt{3}$ and $\alpha=\sqrt{3}$ cases\footnote{The dilatonic solution in the $\alpha=1$ case does not exist below $T_c$.} in AdS$_4$ and the $\alpha=2/\sqrt{6}$ and $\alpha=4/\sqrt{6}$ cases in AdS$_5$. By \eqref{eq:SnF}, the R\'{e}nyi entropies are
\begin{equation}
S_n=\frac{1}{8G}\biggl(1+\frac{1}{n}\biggr)V_\Sigma\,,
\end{equation}
where $V_\Sigma$ is the regulated volume of the hyperbolic space. As a comparison, the R\'{e}nyi entropies for 2D CFTs are
\begin{equation}
S_n^{(2D)}=\frac{c}{6}\biggl(1+\frac{1}{n}\biggr)\log\frac{R}{\delta}\,.
\end{equation}
The $n$ dependence of the R\'{e}nyi entropies for dilatonic systems in supergravity is exactly the same as 2D CFTs, while the R\'{e}nyi entropies calculated from hyperbolic Schwarzschild-AdS black hole contain higher orders in $1/n$.

\subsection{Phase transitions of QFTs in\\ de Sitter space}
\label{sec:dS}
Strongly coupled QFTs in de Sitter space can be studied in terms of the AdS/CFT correspondence. The goal is to find an AdS solution whose boundary is conformal to a de Sitter space in static coordinates:
\begin{equation}
d\hat{s}^2=-(1-H^2\rho^2)\,dt^2+\frac{d\rho^2}{1-H^2\rho^2}+\rho^2d\Omega_{d-2}^2\,,
\end{equation}
where $d\Omega_{d-2}^2$ is the metric for a $(d-2)$-dimensional sphere, and $H$ is the Hubble parameter. The de Sitter space has a temperature given by \cite{Gibbons:1977mu}
\begin{equation}
T_\text{dS}=\frac{H}{2\pi}\,.
\end{equation}

The foliation of de Sitter space gives a solution
\begin{equation}
ds^2 =dr^2 + H^2L^2\tsp \sinh^2\frac{r}{L}\, d\hat{s}^2\,.
\end{equation}
This solution is limited to a special case, in which the temperature $T$ for the QFT and the temperature $T_\text{dS}$ for the de Sitter space are the same. This solution is equivalent to a zero-mass hyperbolic black hole; the general hyperbolic black hole described by~\eqref{eq:Schw} is used for $T\neq T_\text{dS}$ \cite{Marolf:2010tg}. However, there is no phase transition even in this general case \cite{Emparan:1998he,Birmingham:1998nr}.

Here we have a more general class of neutral hyperbolic black holes. The solution of the hyperbolic black hole in AdS$_4$ is~\eqref{eq:hbh} and \eqref{eq:fsol4-1}. After a coordinate transformation by $H\rho=\tanh\theta$, the hyperbolic space $d\tsp\Sigma_2^2$ can be written as\footnote{The static patch of dS$_3$ is conformal to the Lorentzian hyperbolic cylinder $\mathbb{R}\times\mathbb{H}^2$ \cite{Marolf:2010tg}.}
\begin{equation}
d\tsp\Sigma_{2}^2=d\theta^2+\sinh^2\theta\tsp d\varphi^2=\frac{H^2d\rho^2}{(1-H^2 \rho^2)^2} + \frac{H^2\rho^2}{1-H^2\rho^2}\, d\varphi^2\,.\label{eq:coord-change-4}
\end{equation}
By substituting~\eqref{eq:coord-change-4} to~\eqref{eq:hbh}, the hyperbolic black hole in AdS$_4$ can be written as
\begin{multline}
ds^2 = \frac{H^2U(r)}{1-H^2\rho^2} \, \biggl(-\frac{f(r)}{H^2U(r)}\, (1 -H^2\rho^2)\tsp dt^2\\
 + \frac{d\rho^2}{1-H^2\rho^2} + \rho^2\tsp d\varphi^2 \biggr) + \frac{dr^2}{f(r)}\,.
\label{dS-AdS4}
\end{multline}
The conformal boundary is at $r\to\infty$, where $f(r)\to r^2/L^2$ and $U(r)\to r^2$. The QFT lives on the AdS boundary, which is conformal to
\begin{equation}
ds_\partial^2=-\frac{1}{H^2L^2}(1-H^2\rho^2)\,dt^2+\frac{d\rho^2}{1-H^2\rho^2}+\rho^2d\varphi^2\,.\label{eq:bry-dS3}
\end{equation}
According to Sec.~\ref{sec:ads4}, there are phase transitions between hyperbolic black holes with and without scalar hair. Therefore, there will be phase transitions of QFTs in de Sitter space~\eqref{eq:bry-dS3}.

In the case of AdS$_5$, the AdS boundary of the spacetime is conformal to dS$_4$, which can be made explicit by the following coordinates:
\begin{multline}
ds^2 = \frac{H^2U(r)}{1-H^2\rho^2} \, \biggl(-\frac{f(r)}{H^2U(r)}\, (1 -H^2\rho^2)\tsp dt^2\\
 + \frac{d\rho^2}{1-H^2\rho^2} + \rho^2\tsp d\Omega_2^2 \biggr) + \frac{dr^2}{g(r)}\,.
\label{dS-AdS5}
\end{multline}

\section{C-metric solution as a generalization}
\label{sec:C-metric}

When a black hole solution is available, we can try to find a C-metric solution as a generalization with one more parameter. We obtain the C-metric solution for the hyperbolic black holes with scalar hair. As a comparison, the C-metric without scalar hair in Ricci-flat spacetime is
\begin{equation}
ds^2=\frac{1}{a^2(x-y)^2}\biggl(G(y)dt^2-\frac{dy^2}{G(y)}
+\frac{dx^2}{G(x)}+G(x)d\varphi^2\biggr),
\end{equation}
where
\begin{equation}
G(\xi)=1-k\xi^2-2am\xi^3\,.
\end{equation}
The $a\to 0$ limit of this solution gives the Schwarzschild solution. The C-metric solution with a cosmological constant and different $k$ was given in \cite{Mann:1996gj}.

The C-metric solution for an EMD system with $V(\phi)=0$ was given in \cite{Dowker:1993bt}. The C-metric solution for the EMD system with $V(\phi)$ as~\eqref{eq:potential4} in the $k=1$ case was given in \cite{Lu:2014ida}. A charged C-metric is significantly more sophisticated than a neutral one, and thus it is desirable to analyze a neutral C-metric with scalar hair before the charged one.

We keep the gauge field and the general $k=0, \pm 1$ in the following solutions. Only in the $k=-1$ case can we obtain a neutral black hole with scalar hair in the $a\to 0$ limit. The C-metric solution to the system~\eqref{eq:action4} with~\eqref{eq:potential4} is
\begin{multline}
ds^2=\frac{1}{a^2(x-y)^2}\Biggl[h_x^\frac{2\alpha^2}{1+\alpha^2}\left(-F(y)d\tau^2+\frac{dy^2}{F(y)}\right)\\
+h_y^\frac{2\alpha^2}{1+\alpha^2}\left(\frac{dx^2}{G(x)}+G(x)d\varphi^2\right)\Biggr],
\end{multline}
with
\begin{align}
& F =-(1-\gamma y-ky^2-acy^3)h_y^\frac{1-\alpha^2}{1+\alpha^2}+\frac{1}{a^2L^2}h_y^\frac{2\alpha^2}{1+\alpha^2},\nonumber\\
& G =(1-\gamma x-kx^2-acx^3)h_x^\frac{1-\alpha^2}{1+\alpha^2},\nonumber\\
& h_y =1+aby\,,\qquad h_x =1+abx\,,\\
& A =2\sqrt{\frac{bc}{1+\alpha^2}}\,y\tsp dt\,,\qquad e^{\alpha\phi}=\left(\frac{h_y}{h_x}\right)^\frac{2\alpha^2}{1+\alpha^2},\nonumber
\end{align}
where $\gamma$ is an arbitrary constant as a gauge choice. There are two neutral limits: one is $b=0$, in which the dilaton field vanishes; the other is $c=0$. When we take $c=0$ and $k=-1$, we obtain the C-metric generalization of the solution in Sec.~\ref{sec:two}.

After a coordinate transformation \cite{Griffiths:2006tk},
\begin{equation}
y=-\frac{1}{ar}\,,\qquad \tau=at\,,
\end{equation}
the above solution can be written as
\begin{multline}
ds^2=\frac{1}{(1+arx)^2}\left[h_x^\frac{2\alpha^2}{1+\alpha^2}\left(-F(r)dt^2+\frac{dr^2}{F(r)}\right)\right.\\
+\left.r^2h_r^\frac{2\alpha^2}{1+\alpha^2}\left(\frac{dx^2}{G(x)}+G(x)d\varphi^2\right)\right],
\end{multline}
with
\begin{align}
& F=\left(k-\frac{c}{r}-\gamma ar-a^2r^2\right)h_r^\frac{1-\alpha^2}{1+\alpha^2}+\frac{r^2}{L^2}h_r^\frac{2\alpha^2}{1+\alpha^2},\nonumber\\
& G=(1-\gamma x-kx^2-acx^3)h_x^\frac{1-\alpha^2}{1+\alpha^2},\nonumber\\
& h_r =1-\frac{b}{r}\,,\qquad h_x =1+abx\,,\\
& A =2\sqrt{\frac{bc}{1+\alpha^2}}\left(\frac{1}{r_h}-\frac{1}{r}\right)dt\,,\qquad e^{\alpha\phi}=\left(\frac{h_r}{h_x}\right)^\frac{2\alpha^2}{1+\alpha^2}.\nonumber
\end{align}
The $a\to 0$ limit of this solution is explicitly the same as~\eqref{eq:sol4}--\eqref{eq:fsol4}.

We find that there are two cosmic branes with nonconstant tension at $x=0$ and $y=0$, respectively, provided that $\gamma$ is
\begin{equation}
\gamma=\frac{1-3\alpha^2}{1+\alpha^2}ab\,.
\end{equation}
The extrinsic curvature and the induced metric of the hypersurface $x=0$ satisfy (assuming $y<0$)
\begin{equation}
K_{\mu\nu}=\lambda_yh_{\mu\nu},\qquad \lambda_y=-\frac{a\,[1+(1+aby)\alpha^2]}{(1+\alpha^2)(1+a b y)^{\frac{\alpha ^2}{1+\alpha ^2}}}\,.
\end{equation}
The brane tension is
\begin{equation}
\mathcal{T}=-\frac{\lambda_y}{2\pi G_4}\,,
\end{equation}
where $G_4$ is the four-dimensional Newton's constant. When $\alpha=0$ \cite{Emparan:1999wa,Emparan:1999fd}, we have $\lambda_y=-a$. Similarly, the extrinsic curvature and the induced metric of the hypersurface $y=0$ satisfy (assuming $x>0$)
\begin{equation}
K_{\mu\nu}=\lambda_xh_{\mu\nu},\qquad \lambda_x=\frac{\sqrt{1-a^2L^2}\,[1+(1+abx)\alpha^2]}{L(1+\alpha^2)(1+a b x)^{\frac{\alpha ^2}{1+\alpha ^2}}}.
\end{equation}

\section{Discussion}
We have found a class of hyperbolic black holes with scalar hair in AdS space, by taking a particular limit of an EMD system. For the spherical and planar black holes in the EMD solution, the same type of limit does not give a black hole. The main conclusions are summarized as follows:

\begin{itemize}
\item For the Einstein-scalar system we consider, there is an analytic solution for hyperbolic black holes with scalar hair. The system is obtained by taking a neutral limit of an EMD system whose special cases include maximal gauged supergravities, while the dilaton field is kept nontrivial.

\item There are phase transitions between the hyperbolic black hole with scalar hair and the hyperbolic Schwarzschild-AdS black hole. Phase transitions can be zeroth or third order.

\item By holography, the system we study describes (i) phase transitions of R\'{e}nyi entropies and (ii) phase transitions of QFTs in de Sitter space.

\item We give a C-metric solution as a generalization. This neutral C-metric is less complicated than the full EMD solution, while the dilaton field is nontrivial.

\item We propose two constraints for Einstein-scalar systems. Consequently, the potential of the scalar field is highly restricted, and analytic solutions are available. See Appendix~\ref{sec:constraint}.
\end{itemize}

The following topics need further investigation: (i) the special cases of $\alpha=1/\sqrt{3}$, $1$, $\sqrt{3}$ for the charged hyperbolic black holes, (ii) the dual CFT of the hyperbolic black holes \cite{Huang:2014gca}, (iii) the relation to QFTs in cosmological backgrounds \cite{Erdmenger:2011sy}, (iv) the black funnels and droplets \cite{Hubeny:2009ru,Hubeny:2009kz} from the C-metric solutions.

\acknowledgments
I thank Steven Gubser, Christopher Herzog, Elias Kiritsis, Li Li, Yan Liu, Hong L\"{u}, and Shubho Roy for helpful discussions. This work was supported in part by the NSF of China under Grant No. 11905298 and the 100 Talents Program of Sun Yat-sen University under Grant No. 74130-18841203.

\appendix
\section{Holographic renormalization}
\label{sec:HR}
We closely follow \cite{Caldarelli:2016nni} to calculate the mass of the black holes in Einstein-scalar systems by holographic renormalization. The boundary condition for the scalar field corresponds to a multi-trace deformation in the dual CFT.

In the Fefferman-Graham gauge, the AdS$_4$ metric is written in the form
\begin{equation}
ds^2=\frac{L^2}{z^2}(dz^2+g_{ij}(x,z)dx^i dx^j)\,,
\label{eq:FG}
\end{equation}
where $z$ is the FG radial coordinate, and $g_{ij}$ is a three-dimensional metric, which raises/lowers the $i, j$ indexes. The asymptotic expansions of the metric and scalar field are
\begin{align}
g_{ij}(x,z) &=g_{(0)ij}+zg_{(1)ij}+z^2g_{(2)ij}+\cdots,\\
\phi(x,z) &=z^{\Delta_-}\varphi(x,z)=z^{\Delta_-}(\varphi_{(0)}+z\varphi_{(1)}+\cdots),
\end{align}
where $\Delta_-=1$ for the solution we consider. The boundary condition for a multi-trace deformation is specified starting with the alternative quantization. The single-trace source is written as 
\begin{equation}
J_\mathcal{F}=-L^2\varphi_{(1)}-\mathcal{F}'(\varphi_{(0)})\,,
\end{equation}
where $\mathcal{F}(\varphi_{(0)})$ is a polynomial, and $J_\mathcal{F}=0$ specifies the sourceless condition. With a general boundary condition for the scalar field, we need to add an additional finite boundary term $S_\mathcal{F}$ to the renormalized on-shell action:
\begin{equation}
S_\text{ren}=\lim_{\epsilon\to 0}(S_\text{bulk}+S_\text{GH}+S_\text{ct}+S_\mathcal{F})\,,
\label{eq:Sren}
\end{equation}
where $S_\text{bulk}$ is the action \eqref{eq:action4} with $r$ being integrated from the horizon $r_h$ to the cutoff $z=\epsilon$, and
\begin{align}
S_\text{GH} &=\int_{z=\epsilon} d^3x\sqrt{-\gamma}\,2K\,,\\
S_\text{ct} &=-\int_{z=\epsilon} d^3\sqrt{-\gamma}\Bigl(\frac{4}{L}+LR[\gamma]+\frac{1}{2L}\phi^2\Bigr),\\
S_\mathcal{F} &=\int_{z=\epsilon} d^3x\sqrt{-g_{(0)}}\Bigl(J_\mathcal{F}\varphi_{(0)}+\mathcal{F}(\varphi_{(0)})\Bigr).
\label{eq:SF}
\end{align}
Its variation is
\begin{equation}
\delta S_\text{ren}=\int d^3x\sqrt{-g_{(0)}}\Bigl(\frac{1}{2}\langle\mathcal{T}^{ij}\rangle\delta g_{(0)}^{ij}+\langle O_{\Delta_-}\rangle\delta J_\mathcal{F}\Bigr).
\end{equation}
The boundary stress tensor is given by \cite{Caldarelli:2016nni}
\begin{equation}
\langle\mathcal{T}^{ij}\rangle=3L^2g_{(3)}^{ij}+\Bigl(\mathcal{F}(\varphi_{(0)})-\varphi_{(0)}\mathcal{F}'(\varphi_{(0)})\Bigr)g_{(0)}^{ij}\,.
\label{eq:Tij}
\end{equation}

For the solution~\eqref{eq:hbh} with \eqref{eq:fsol4-1}, the relation between the coordinates $r$ and $z$ is
\begin{align}
r=&\frac{L^2}{z}+\frac{\alpha ^2 b}{1+\alpha ^2}+\biggl(\frac{1}{4}+\frac{\alpha ^2 b^2}{4 L^2 (1+\alpha ^2)^2}\biggr) z\nonumber\\
&+\biggl(\frac{c}{6 L^2}-\frac{(1-\alpha ^2) b}{6 L^2 (1+\alpha ^2)}+\frac{\alpha ^2 (1-\alpha ^2) b^3}{9 L^4 (1+\alpha ^2)^3}\biggr) z^2+\cdots.
\end{align}
The metric and the scalar field in the FG expansion are
\begin{align}
& g_{ij}=\left(\begin{array}{ccc}
-1 & 0 & 0\\
0 & L^2 & 0\\
0 & 0 & L^2s_\theta^2
\end{array}\right)\nonumber\\
& \quad +z^2\left(\begin{array}{ccccc}
\frac{1}{2L^2}+\frac{\alpha ^2 b^2}{2 L^4 (1+\alpha ^2)^2} \span\span & 0 & 0\\
0 & \span \frac{1}{2}-\frac{\alpha ^2 b^2}{2 L^2 (1+\alpha ^2)^2} \span & 0\\
0 & 0 & \span\span \Bigl(\frac{1}{2}-\frac{\alpha ^2 b^2}{2 L^2 (1+\alpha ^2)^2}\Bigr)s_\theta^2
\end{array}\right)\nonumber\\
& \quad +z^3\left(\begin{array}{ccccc}
\frac{2c}{3L^4}-\frac{2(1-\alpha ^2) b}{3 L^4 (1+\alpha ^2)}+\frac{4\alpha ^2 (1-\alpha ^2) b^3}{9 L^6 (1+\alpha ^2)^3} \span\span & 0 & 0\\
0 & \span \frac{c}{3L^2}-\frac{(1-\alpha ^2) b}{3 L^2 (1+\alpha ^2)}-\frac{4\alpha ^2 (1-\alpha ^2) b^3}{9 L^4 (1+\alpha ^2)^3}  \span & 0\\
0 & 0 & \span\span \Bigl(\frac{c}{3L^2}-\frac{(1-\alpha ^2) b}{3 L^2 (1+\alpha ^2)}-\frac{4\alpha ^2 (1-\alpha ^2) b^3}{9 L^4 (1+\alpha ^2)^3}\Bigr)s_\theta^2
\end{array}\right)\nonumber\\
& \quad +\cdots,\\
& \phi=-\frac{2\alpha b}{L^2(1+\alpha^2)}z-\frac{\alpha (1-\alpha^2) b^2}{L^4(1+\alpha^2)^2}z^2+\cdots,\label{eq:phi-FG}
\end{align}
where $s_\theta\equiv\sinh\theta$. From \eqref{eq:phi-FG}, we can read
\begin{equation}
\varphi_{(0)}=-\frac{2\alpha b}{L^2(1+\alpha^2)},\qquad \varphi_{(1)}=-\frac{\alpha (1-\alpha^2) b^2}{L^4(1+\alpha^2)^2}.
\end{equation}

The solution~\eqref{eq:fsol4-1} is compatible with a triple-trace deformation with a marginal coupling $\vartheta$:
\begin{equation}
\mathcal{F}=\frac{1}{3}\vartheta\varphi_{(0)}^3\,,\qquad \vartheta=\frac{1-\alpha^2}{4\alpha}L^2\,.
\end{equation}
This gives an AdS-invariant boundary condition despite the fact that the metric falls off slower than usual \cite{Hertog:2004dr}. The choice of the boundary term~\eqref{eq:SF} ensures that the first law of thermodynamics is unmodified by a scalar charge, consistent with \cite{Faulkner:2010gj}. By \eqref{eq:Tij}, the extra terms of $\mathcal{F}$ cancel the $b^3$ terms in $g_{(3)}^{ij}$, and the boundary stress tensor is traceless. The energy density is given by
\begin{equation}
\varepsilon=L^2\langle\mathcal{T}^{tt}\rangle
\end{equation}
Consequently, the mass of the hyperbolic black hole is given by \eqref{eq:thermo-M}, and the free energy is given by \eqref{eq:thermo-F}. As a crosscheck, directly evaluating the on-shell action \eqref{eq:Sren} gives $S^E_\text{ren}=-iS_\text{ren}=\beta F$ ($\beta=1/T$), which is in agreement with \eqref{eq:thermo-F}.\footnote{Unlike planar black holes, the bulk Lagrangian is not a total derivative for hyperbolic black holes, and thus, the on-shell action inevitably contains an integral from the horizon to the AdS boundary. For more details on the difference between planar and hyperbolic/spherical black holes, see Appendix A of \cite{Bai:2022obp}.}

The solution~\eqref{eq:fsol4-1} is also compatible with a double-trace deformation
\begin{equation}
\mathcal{F}=\frac{1}{2}\vartheta\varphi_{(0)}^2\,,\qquad \vartheta=-\frac{1-\alpha^2}{2(1+\alpha^2)}b\,.
\end{equation}
The mass is
\begin{equation}
M=\frac{V_\Sigma}{8\pi G}\left(-\frac{1-\alpha^2}{1+\alpha^2}b+\frac{\alpha^2(1-\alpha^2)}{6(1+\alpha^2)^3}b^3\right).
\end{equation}
However, we cannot change the temperature or entropy for the analytic solutions if we fix $\vartheta$, since $b$ and $r_h$ are related by~\eqref{eq:brh}. Instead, we can obtain numerical solutions of hyperbolic black holes at different temperatures by fixing the double-trace deformation parameter.

\section{Special cases of STU supergravity}
\label{sec:STU}
In STU supergravities, there are U$(1)^4$ gauge fields in AdS$_4$, U$(1)^3$ gauge fields in AdS$_5$, and U$(1)^2$ gauge fields in AdS$_7$ \cite{Cvetic:1999xp}. Special cases of them can be reduced to EMD systems. They are 1-charge, 2-charge, and 3-charge black holes in AdS$_4$; 1-charge and 2-charge black holes in AdS$_5$; and 1-charge black hole in AdS$_7$.

The AdS$_4$ Lagrangian is
\begin{multline}
\mathcal{L}=R-\frac{1}{2}(\partial\vec{\phi})^2+8g^2(\cosh\phi_1+\cosh\phi_2+\cosh\phi_3)\\
-\frac{1}{4}\sum_{i=1}^4 e^{\vec{a}_i\cdot\vec{\phi}}(F^i_{(2)})^2\,,
\end{multline}
where $\vec{\phi}=(\phi_1, \phi_2, \phi_3)$, $\vec{a}_1=(1, 1, 1)$, $\vec{a}_2=(1, -1, -1)$, $\vec{a}_3=(-1, 1, -1)$, and $\vec{a}_4=(-1, -1, 1)$. More details can be found in \cite{Cvetic:1999xp}. The solution is given by \cite{Duff:1999gh,Sabra:1999ux}
\begin{align}
& ds^2=-(H_1H_2H_3H_4)^{-1/2}fdt^2\nonumber\\
& \hspace{30pt}+(H_1H_2H_3H_4)^{1/2}(f^{-1}d\bar{r}^2+\bar{r}^2d\tsp\Sigma_{2,k}^2)\,,\\
& X_i=H_i^{-1}(H_1H_2H_3H_4)^{1/4}\,,\\
& A_{(1)}^i=\sqrt{k}(1-H_i^{-1})\coth\beta_i\tsp dt\,,
\end{align}
with $X_i=e^{-\frac{1}{2}\vec{a}_i\cdot\vec{\phi}}$, and
\begin{equation}
f=k-\frac{\mu}{\bar{r}}+\frac{4}{L^2}\bar{r}^2(H_1H_2H_3H_4)\,,\qquad H_i=1+\frac{\mu\sinh^2\beta_i}{k\bar{r}}\,.
\end{equation}

The following special cases are obtained when some of the U$(1)^4$ charges are the same, and others are zero:
\begin{itemize}
\item 1-charge black hole ($\alpha=\sqrt{3}$): $H_1=H$, $H_2=H_3=H_4=1$. The Lagrangian is
\begin{equation}
\mathcal{L}=R-\frac{1}{2}(\partial\phi)^2+\frac{6}{L^2}\cosh\frac{\phi}{\sqrt{3}}-\frac{1}{4}e^{-\sqrt{3}\phi}F^2\,.
\end{equation}

\item 2-charge black hole ($\alpha=1$): $H_1=H_2=H$, $H_3=H_4=1$. The Lagrangian is
\begin{equation}
\mathcal{L}=R-\frac{1}{2}(\partial\phi)^2+\frac{2}{L^2}(\cosh\phi+2)-\frac{1}{4}e^{-\phi}F^2\,.
\end{equation}

\item 3-charge black hole ($\alpha=1/\sqrt{3}$): $H_1=H_2=H_3=H$, $H_4=1$. The Lagrangian is
\begin{equation}
\mathcal{L}=R-\frac{1}{2}(\partial\phi)^2+\frac{6}{L^2}\cosh\frac{\phi}{\sqrt{3}}-\frac{1}{4}e^{-\frac{1}{\sqrt{3}}\phi}F^2\,.
\end{equation}

\item 4-charge black hole ($\alpha=0$): $H_1=H_2=H_3=H_4=H$. This is the RN-AdS$_4$ black hole.
\end{itemize}

The AdS$_5$ Lagrangian is
\begin{equation}
\mathcal{L}=R-\frac{1}{2}(\partial\vec{\varphi})^2+4g^2\sum_i X_i^{-1}-\frac{1}{4}\sum_{i=1}^4 X_i^{-2} (F^i_{(2)})^2\,.
\end{equation}
The solution is \cite{Behrndt:1998jd}
\begin{align}
& ds^2=-(H_1H_2H_3)^{-2/3}fdt^2\nonumber\\
&\hspace{30pt}+(H_1H_2H_3)^{1/3}(f^{-1}d\bar{r}^2+\bar{r}^2d\tsp\Sigma_{3,k}^2)\,,\\
& X_i=H_i^{-1}(H_1H_2H_3)^{1/3}\,,\\
& A_{(1)}^i=\sqrt{k}(1-H_i^{-1})\coth\beta_idt\,,
\end{align}
with
\begin{equation}
f=k-\frac{\mu}{\bar{r}}+\frac{4}{L^2}\bar{r}^2(H_1H_2H_3)\,,\qquad H_i=1+\frac{\mu\sinh^2\beta_i}{k\bar{r}}\,.
\end{equation}

The following special cases are obtained when some of the U$(1)^3$ charges are the same, and others are zero:
\begin{itemize}
\item 1-charge black hole ($\alpha=4/\sqrt{6}$): $H_1=H$, $H_2=H_3=1$. The Lagrangian is
\begin{equation}
\mathcal{L}=R-\frac{1}{2}(\partial\phi)^2+\frac{4}{L^2}(2e^{\frac{1}{\sqrt{6}}\phi}+e^{-\frac{2}{\sqrt{6}}\phi})-\frac{1}{4}e^{-\frac{4}{\sqrt{6}}\phi}F^2\,.
\end{equation}
\item 2-charge black hole ($\alpha=2/\sqrt{6}$): $H_1=H_2=H$, $H_3=1$. The Lagrangian is
\begin{equation}
\mathcal{L}=R-\frac{1}{2}(\partial\phi)^2+\frac{4}{L^2}(2e^{-\frac{1}{\sqrt{6}}\phi}+e^{\frac{2}{\sqrt{6}}\phi})-\frac{1}{4}e^{-\frac{2}{\sqrt{6}}\phi}F^2\,.
\end{equation}
\item 3-charge black hole ($\alpha=0$): $H_1=H_2=H_3=1$. This is the RN-AdS$_5$ black hole.
\end{itemize}

In the AdS$_7$ case, a similar analysis can be done. There will be 1-charge black hole and 2-charge black hole, and the latter is the RN-AdS$_7$ black hole.

If we set $\mu=0$ in the above solutions, both gauge fields and dilaton fields will vanish. However, in the above EMD systems, it is possible to make the gauge field vanish while keeping the dilaton field nontrivial in the $k=-1$ case, which is a key observation made in this paper. To see this explicitly, we can replace $\beta$ with $i\beta$ and make the following coordinate transformation:
\begin{equation}
r=\bar{r}+\mu\sin^2\beta.
\end{equation}

\section{Constraints on Einstein-scalar systems}
\label{sec:constraint}
We find that the Einstein-scalar system is significantly simplified under some reasonable constraints. We consider the AdS$_4$ spacetime, and the generalization to higher-dimensional cases is straightforward. The action is
\begin{equation}
S=\int d^{4}x\,\sqrt{-g}\left(R-\frac{1}{2}(\partial\phi)^2-V(\phi)\right),\label{eq:Einstein-scalar}
\end{equation}
where $V(\phi)$ is the potential of the scalar field $\phi$. We consider the following metric ansatz:
\begin{equation}
ds^2=e^{2\mathcal{A}(\bar{r})}(-h(\bar{r})dt^2+d\tsp\Sigma_{2,k}^2)+\frac{e^{2\mathcal{B}(\bar{r})}}{h(\bar{r})}d\bar{r}^2\,,\label{eq:ABh-ansatz-4}
\end{equation}
where $\bar{r}$ is the AdS radial coordinate, and the metric for the 2-dimensional sphere, plane, and hyperbloid can be written as
\begin{equation}
d\tsp\Sigma_{2,k}^2=\frac{dx^2}{1-kx^2}+x^2dy^2\,,
\end{equation}
where $k=1$, $0$, and $-1$, respectively. There are four unknown functions $\mathcal{A}(\bar{r})$, $\mathcal{B}(\bar{r})$, $h(\bar{r})$, and $\phi(\bar{r})$ and one gauge degree of freedom.

We propose the following constraints:
\emph{
(i) The potential $V$ is independent of $k$;
(ii) The function $h$ depends on $k$, and other functions are independent of $k$.}
In other words, for a given $V(\phi)$, the only difference between the $k=0$ solution and the $k\neq 0$ solution is some terms $h^{(k)}$ in $h$. Justification of these constraints includes special cases of STU supergravity in Appendix~\ref{sec:STU}. We draw the following statements.

\emph{
For a given cosmological constant $V(0)=-6/L^2$, the general potential satisfying the above constraints is a two-parameter family of the potential given by
\begin{equation}
V_{\alpha,\beta}(\phi)=V_\alpha(\phi)+\beta\left(V_\alpha(\phi)-V_{-\alpha}(\phi)\right),
\label{eq:potential4ab}
\end{equation}
where $\alpha$ and $\beta$ are parameters, and $V_\alpha(\phi)$ is a one-parameter family of the potential given by
\begin{multline}
V(\phi)=-\frac{2}{(1+\alpha^2)^2L^2}\Bigl[\alpha^2(3\alpha^2-1)e^{-\phi/\alpha}\\
+8\alpha^2e^{(\alpha-1/\alpha)\phi/2}+(3-\alpha^2)e^{\alpha\phi}\Bigr].\label{eq:potential4a}
\end{multline}
For the one-parameter family of the potential $V_\alpha(\phi)$, the general solution under the above constraints for the hyperbolic black hole is obtained as~\eqref{eq:hbh} and \eqref{eq:fsol4-1} in Sec.~\ref{sec:two}, and solution under the above constraints for the spherical or planar black hole does not exist.
}

A brief proof is as follows. The Einstein-scalar system has a weak form of integrability, which was presented in \cite{Gubser:2008ny} and \cite{Li:2011hp,Cai:2012xh}. We will review this procedure and give more insights that enable us to find a ``privileged'' potential. Consequently, we find a way to derive the potential~\eqref{eq:potential4a} by relating the $k=0$ and $k\neq 0$ solutions.

Equations of motion are obtained by the action~\eqref{eq:Einstein-scalar} with the metric ansatz~\eqref{eq:ABh-ansatz-4}. The Einstein's equation gives
\begin{gather}
\mathcal{A}'\mathcal{B}'=\frac{1}{4}\phi'^2+\mathcal{A}''\,,\label{eq:solve-B}\\
(e^{3\mathcal{A}-\mathcal{B}}h')'+2e^{\mathcal{A}+\mathcal{B}}k=0\,.\label{eq:solve-h}
\end{gather}
The first equation comes from eliminating the potential $V$ from $G_{tt}=\frac{1}{2}T_{tt}$ and $G_{\bar{r}\bar{r}}=\frac{1}{2}T_{\bar{r}\bar{r}}$. The second equation comes from eliminating the potential $V$ from $G_{\bar{r}\bar{r}}=\frac{1}{2}T_{\bar{r}\bar{r}}$ and $G_{xx}=\frac{1}{2}T_{xx}$. Solving $V$ from $G_{\bar{r}\bar{r}}=\frac{1}{2}T_{\bar{r}\bar{r}}$ gives
\begin{equation}
V=\frac{1}{2}e^{-2\mathcal{B}}(\phi'^2h-12\mathcal{A}'^2h-4\mathcal{A}'h')+2ke^{-2\mathcal{A}}\,.\label{eq:solve-V}
\end{equation}

There is a gauge freedom, which is fixed by $\phi=\bar{r}$. Other equations can be derived from~\eqref{eq:solve-B}, \eqref{eq:solve-h}, and \eqref{eq:solve-V}. Starting with a given $A(\bar{r})$, we can obtain $V(\bar{r})$ in the following way
\begin{equation}
\mathcal{A}\xrightarrow{\eqref{eq:solve-B}} \mathcal{B}\xrightarrow{\eqref{eq:solve-h}} h\xrightarrow{\eqref{eq:solve-V}} V\,.\label{eq:ABhV}
\end{equation}
The function $\mathcal{A}(\bar{r})$ plays the role of a generating function. Finally, replacing the function $V(\bar{r})$ with $V(\phi)$ gives the potential. This method can be generalized to EMD systems in a straightforward way. Only careful choices of $\mathcal{A}$ can we obtain a relatively simple $V(\phi)$. Related methods are used in \cite{Anabalon:2012ta,Feng:2013tza}, in which the potential can be generated by choices of either the metric or the scalar field.

Let $h^{(0)}$ be the solution of $h$ at $k=0$, and we can decompose $h$ into two parts:
\begin{equation}
h(\bar{r})=h^{(0)}+h^{(k)}\,.\label{eq:hk}
\end{equation}
The potential $V$ solved by~\eqref{eq:solve-V} apparently depends on $k$. Constraint (i) requires that the terms dependent on $k$ must cancel:
\begin{equation}
V=V^{(0)}+V^{(k)},\qquad V^{(k)}=0\,.
\end{equation}
If $\mathcal{A}$, $\mathcal{B}$, and $h^{(0)}$ satisfy the equations of motion with $V=V^{(0)}$ in the $k=0$ case, then $\mathcal{A}$, $\mathcal{B}$, and $h^{(k)}$  satisfy the equations of motion with $V=0$ in the $k=1$ case. The key point is that these two cases share the same generating function $\mathcal{A}$, and the latter one is simpler to solve.
A solution with $V=0$ and $k=1$ has been given in~\cite{Garfinkle:1990qj}.

The solution for $h$ from~\eqref{eq:solve-h} is given by
\begin{equation}
h=\int e^{-3\mathcal{A}+\mathcal{B}}\left(-2k\int e^{\mathcal{A}+\mathcal{B}}d\bar{r}+C_2\right)d\bar{r}+C_1\,,
\end{equation}
where the $C_1=C_2=0$ solution is the solution for the system with $V=0$. To satisfy constraint (i), $h$ can only linearly depend on $k$. If we take $C_1=1$ and $C_2\propto k$, we obtain the potential $V$ as~\eqref{eq:potential4a}. Let $V_\alpha(\phi)$ be the potential~\eqref{eq:potential4a} with parameter $\alpha$, the general solution for the potential is
\begin{equation}
V(\phi)=\beta_1 V_\alpha(\phi)+\beta_2 V_{-\alpha}(\phi)\,,
\end{equation}
where $\beta_1$ and $\beta_2$ are constants. This potential can be rewritten as~\eqref{eq:potential4ab}, where the terms proportional to $\beta$ do not change the cosmological constant and the mass of the scalar field. A nonzero $\beta$ will give a cumbersome solution of $h$, in which $C_1=1$, and $C_2$ is related to $\beta$. This potential is expected to be consistent with \cite{Feng:2013tza}, in which \eqref{eq:soln-phi} below was used as an assumption.

Since~\eqref{eq:solve-B} and \eqref{eq:solve-h} are independent of the cosmological constant, the two integration constants $C_1$ and $C_2$ are related to the cosmological constant. A shortcut to obtain the potential~\eqref{eq:potential4a} is as follows. Start with a solution with $V(\phi)=0$ and $k=1$, and then use the procedure~\eqref{eq:ABhV} with $h=1$ and $k=0$.

To solve the equations of motion with $V=0$ and $k=1$, it is more convenient to choose another gauge, $\mathcal{B}=-\mathcal{A}$. The metric ansatz is~\eqref{eq:sol4}. Consider Einstein's equation with left-hand side being $R_{\mu\nu}$, and we take the following procedure:
\begin{itemize}
\item[(a)] From the $tt$ and $\theta\theta$ components, a simple equation is obtained $(fU)''=2$. So $fU$ is a second-order polynomial of $r$, and we can parameterize it as
\begin{equation}
fU=(r-r_1)(r-r_2)\,.\label{eq:soln-fU}
\end{equation}

\item[(b)] The $tt$ component gives $(f'U)'=0$. The general solution of $f$ is
\begin{equation}
f=f_0\left(\frac{r-r_1}{r-r_2}\right)^{\nu_1},\label{eq:soln-f}
\end{equation}
where $f_0$ and $\nu_1$ are integration constants. At the AdS boundary $r\to\infty$, $f=1$ gives $f_0=1$.

\item[(c)] The equation of motion for the scalar field is $(fU\phi')'=0$. The general solution of $\phi$ is
\begin{equation}
e^\phi=e^{\phi_0}\left(\frac{r-r_1}{r-r_2}\right)^{\nu_2},\label{eq:soln-phi}
\end{equation}
where $\phi_0$ and $\nu_2$ are integration constants. At the AdS boundary $r\to\infty$, $\phi=0$ gives $\phi_0=0$.

\item[(d)] Equations~\eqref{eq:soln-fU}--$\eqref{eq:soln-phi}$ are solution to the equations of motion, provided that the condition $\nu_1^2+\nu_2^2=1$ is satisfied.

\item[(e)] The coordinate $\bar{r}$ in the $\phi=\bar{r}$ gauge and the coordinate $r$ in the $\mathcal{A}=-\mathcal{B}$ gauge are related by $\phi(r)=\bar{r}$. By comparing~\eqref{eq:sol4} and \eqref{eq:ABh-ansatz-4}, we have
\begin{equation}
e^\mathcal{A}=(r_1-r_2)\Bigl(e^{\frac{1+\nu_1}{2\nu_2}\bar{r}}-e^{-\frac{1-\nu_1}{2\nu_2}\bar{r}}\Bigr)^{-1}.
\end{equation}
After taking $r_1=b$, $r_2=0$, and $\alpha=(1-\nu_1)/\nu_2$, we obtain the generating function for the potential~\eqref{eq:potential4a} as
\begin{equation}
e^\mathcal{A}=b\left(e^{\frac{1}{2\alpha}\bar{r}}-e^{-\frac{\alpha}{2}\bar{r}}\right)^{-1}.
\end{equation}
\end{itemize}

For higher-dimensional cases, the potential~\eqref{eq:potentiald} can be generated by the same procedure. We replace $d\tsp\Sigma_{2,k}^2$ with $d\tsp\Sigma_{d-1,k}^2$ in \eqref{eq:ABh-ansatz-4}, and the generating function is
\begin{equation}
e^\mathcal{A}=b\left(e^{\frac{d-2}{(d-1)\alpha}\bar{r}}-e^{-\frac{\alpha}{2}\bar{r}}\right)^{-\frac{1}{d-2}}.
\end{equation}

\section{Properties of the planar solution ($k=0$)}
\label{sec:planar}
The hyperbolic black hole solution is related to a planar black hole in the following way. The action with $(d-1)$ axion (massless scalar) fields $\chi_i$ is
\begin{multline}
S=\int d^{d+1}x\sqrt{-g}\Biggl(R-\frac14 e^{-\alpha\phi}F^2-\frac12(\partial\phi)^2-V(\phi)\\
-\frac{1}{2}\sum_{i=1}^{d-1}(\partial\chi_i)^2\Biggr),\label{eq:axionsd}
\end{multline}
where $\chi_i=\kappa x_i$ satisfies the equation of motion of $\chi_i$. This system was used as a simple way to introduce momentum dissipation, since $\kappa x_i$ breaks the translation symmetry \cite{Andrade:2013gsa,Gouteraux:2014hca}. As pointed out in~\cite{Gouteraux:2014hca}, for the potential~\eqref{eq:potentiald}, the solution to the functions with the metric ansatz~\eqref{eq:ansatzd-1} in the $k=0$ case is the same as that of a hyperbolic black hole without the axions $\chi_i$.

For an arbitrary potential $V(\phi)$ with the metric ansatz
\begin{equation}
ds^2=e^{2\mathcal{A}(\bar{r})}(-h(\bar{r})dt^2+d\tsp\Sigma_{d-1,k}^2)+\frac{e^{2\mathcal{B}(\bar{r})}}{h(\bar{r})}d\bar{r}^2\,,\label{eq:ABh-ansatz-d}
\end{equation}
we observe that the equations of motion for a planar black hole with axions are the same as the equations of motion for a hyperbolic black hole without axions, provided that we make the identification
\begin{equation}
k=-\frac{1}{2(d-2)}\kappa^2\,.
\end{equation}
Here $\sqrt{|k|}$ is the inverse curvature radius of the hyperbolic space, and we set it to $1$ previously. The neutral planar black holes with momentum dissipation was studied in \cite{Ren:2021rhx}.

The planar solution without axions has an intriguing (and peculiar) property: for $\alpha>(d-2)\sqrt{\frac{2}{d(d-1)}}$, the gauge field is automatically eliminated from the EMD system in the extremal limit $r_h\to b$. In this range of $\alpha$, the gauge field is proportional to a positive power of $(r_h-b)$. Examples include 1-charge black hole in AdS$_5$ \cite{DeWolfe:2012uv}; 1-charge and 2-charge black holes in AdS$_4$ \cite{Kiritsis:2015oxa}.

Consider the AdS$_4$ solution. Starting with the charged planar black hole, we have $b>0$ and $c>0$. The condition $f(r_h)=0$ gives
\begin{equation}
c=\frac{1}{L^2}r_h^\frac{4}{1+\alpha^2}(r_h-b)^\frac{3\alpha^2-1}{1+\alpha^2}.
\end{equation}
When $\alpha>1/\sqrt{3}$, we have $c=0$ when we take the extremal limit $r_h\to b$. Consequently, the gauge field vanishes since $A_t\propto\sqrt{bc}$. The extremal solution is
\begin{align}
ds^2 &=f\tsp (-dt^2+d\vec{x}^2)+f^{-1}dr^2,\qquad A=0\,,\nonumber\\
f &=\frac{r^2}{L^2}\left(1-\frac{b}{r}\right)^\frac{2\alpha^2}{1+\alpha^2},
\qquad e^{\alpha\phi}=\left(1-\frac{b}{r}\right)^\frac{2\alpha^2}{1+\alpha^2}.
\end{align}
This is the neutral limit of \eqref{eq:sol4}--\eqref{eq:fsol4} in the $k=0$ case, i.e., the planar counterpart of \eqref{eq:hbh} and \eqref{eq:fsol4-1}. This solution does not have a horizon, and has a spacetime singularity at $r=b$. We expect that this solution can be taken as an extremal limit of a finite temperature solution, when the Gubser criterion \cite{Gubser:2000nd} is satisfied. The finite temperature solution will not satisfy the constraints proposed in Appendix~\ref{sec:constraint}. By a coordinate transformation $r=\bar{r}+b$, the above solution becomes
\begin{align}
ds^2 &=f\tsp (-dt^2+d\vec{x}^2)+f^{-1}d\bar{r}^2,\qquad A=0\,,\nonumber\\
f &=\frac{\bar{r}^2}{L^2}\left(1+\frac{b}{\bar{r}}\right)^\frac{2}{1+\alpha^2},\qquad e^{-\alpha\phi}=\left(1+\frac{b}{\bar{r}}\right)^\frac{2\alpha^2}{1+\alpha^2}.
\end{align}
The spacetime singularity is now at $\bar{r}=0$.

The planar black hole solution to the EMD system~\eqref{eq:actiond} has the following distinctive IR geometries in the extremal case.
\begin{itemize}
\item $0<\alpha<(d-2)\sqrt{\frac{2}{d(d-1)}}$. The IR geometry is AdS$_2\times\mathbb{R}^{d-1}$.
\item $\alpha=(d-2)\sqrt{\frac{2}{d(d-1)}}$. The IR geometry is conformal to AdS$_2\times\mathbb{R}^{d-1}$.
\item $\alpha>(d-2)\sqrt{\frac{2}{d(d-1)}}$. The extremal limit of the EMD system~\eqref{eq:actiond} is the same as an Einstein-scalar system. The IR geometry is a hyperscaling-violating geometry.
\end{itemize}

If we treat the Einstein-scalar (neutral) system as the starting point, the parameter $b$ can be either positive or negative. We draw the following conclusions by analyzing the IR geometry according to \cite{Charmousis:2010zz} (see also \cite{Kiritsis:2015oxa}). If $b<0$, which implies $\alpha\phi>0$, the leading term in $V(\phi)$ in the IR is the last term, and we have
\begin{itemize}
\item $0<\alpha<\sqrt{\frac{2}{d-1}}$. The spectrum is gapless.
\item $\sqrt{\frac{2}{d-1}}\leq\alpha<\sqrt{\frac{2d}{d-1}}$. The extremal geometry is at $T\to\infty$. The spectrum is potentially gapped.
\item $\alpha\geq\sqrt{\frac{2d}{d-1}}$. It violates the Gubser criterion, and thus unacceptable holographically. However, the IR geometry can be changed by introducing extra fields.
\end{itemize}

If $b>0$, which implies $\alpha\phi<0$, the leading term in $V(\phi)$ in the IR is the first term. Furthermore, the potential is invariant under the transformation
\begin{equation}
\alpha\to\frac{2(d-2)}{(d-1)\alpha},\qquad \phi\to -\phi\,.
\end{equation}
We draw the following conclusion for $b>0$:
\begin{itemize}
\item $\alpha>(d-2)\sqrt{\frac{2}{d-1}}$. The spectrum is gapless.
\item $(d-2)\sqrt{\frac{2}{d(d-1)}}<\alpha\leq (d-2)\sqrt{\frac{2}{d-1}}$. The extremal geometry is at $T\to\infty$. The spectrum is potentially gapped.
\item $0<\alpha\leq (d-2)\sqrt{\frac{2}{d(d-1)}}$. It violates the Gubser criterion, and thus unacceptable holographically.
\end{itemize}

\end{document}